\documentclass[%
 reprint,
superscriptaddress,
 amsmath,amssymb,
 aps,
]{revtex4-1}

\usepackage{graphicx}
\usepackage{dcolumn}
\usepackage{mathtools}
\usepackage{bm}
\usepackage{float}
\usepackage{epstopdf}
\usepackage[normalem]{ulem}
\usepackage{colortbl}
\usepackage{hyperref}
\usepackage{ulem}

\usepackage{braket}
\usepackage{bbm}
\usepackage[dvipsnames]{xcolor}
\usepackage{verbatim}
\usepackage{multirow}

\begin{document}

\title{Continuous variable multimode quantum states\\
via symmetric group velocity matching}

\author{V.~Roman-Rodriguez}
\affiliation{Sorbonne Universit\'e, CNRS, LIP6, 4 place Jussieu, F-75005 Paris, France}
\author{B.~Brecht}
\affiliation{Integrated Quantum Optics, Institute for Photonic Quantum Systems (PhoQS), Paderborn University, Warburger Stra{\ss}e 100 33098,
Paderborn, Germany}
\author{S.Kaali}
\affiliation{Laboratoire Kastler Brossel, Sorbonne Universit\'{e}, CNRS, ENS-PSL Research University, Coll\`{e}ge de France, 4 place Jussieu, F-75252 Paris, France}
\author{C.~Silberhorn}
\affiliation{Integrated Quantum Optics, Institute for Photonic Quantum Systems (PhoQS), Paderborn University, Warburger Stra{\ss}e 100 33098,
Paderborn, Germany}
\author{N.~Treps}
\affiliation{Laboratoire Kastler Brossel, Sorbonne Universit\'{e}, CNRS, ENS-PSL Research University, Coll\`{e}ge de France, 4 place Jussieu, F-75252 Paris, France}
\author{E.Diamanti}
\affiliation{Sorbonne Universit\'e, CNRS, LIP6, 4 place Jussieu, F-75005 Paris, France}
\author{V.~Parigi}
\affiliation{Laboratoire Kastler Brossel, Sorbonne Universit\'{e}, CNRS, ENS-PSL Research University, Coll\`{e}ge de France, 4 place Jussieu, F-75252 Paris, France}
\email{vickrfb@gmail.com}

\date{\today}

\begin{abstract}
\noindent Configurable and scalable continuous variable quantum networks for measurement-based quantum information protocols or multipartite quantum communication schemes can be obtained via parametric down conversion (PDC) in non-linear waveguides. In this work, we exploit symmetric group velocity matching (SGVM)  to engineer the properties of the squeezed modes of the PDC. We identify type \emph{II} PDC in a single waveguide as the best suited process, since multiple modes with non-negligible amount of squeezing can be obtained. We explore, for the first time, the waveguide dimensions, usually only set to ensure single-mode guiding, as an additional design parameter ensuring indistinguishability of the signal and idler fields. We investigate here  potassium titanyl phosphate (KTP), which offers SGVM at telecommunications wavelengths, but our approach can be applied to any non-linear material and pump wavelength. This work paves the way towards the engineering of future large-scale quantum networks in the continuous variable regime.
\end{abstract}

\maketitle

\section{Introduction}

Photonic quantum technologies are grounded on the remarkable ability of engineering quantum states of light in different non-linear optical processes \cite{SPOPO,OpticaPaderborn}. Besides the versatility in producing different quantum states, optics has the advantage of offering scalability in the number of systems by exploiting multimode electro-magnetic fields. This is even more relevant when we consider a continuous variable (CV) encoding of quantum information since it is possible to deterministically generate large multimode entangled states,  where entanglement is established between amplitude and phase quadratures of different light modes \cite{SPOPO,Chen14,Yokoyama13,Cai17,Asavanant19}. CV multimode entangled states are resources for measurement based quantum computing  \cite{Menicucci06,Gu09},  quantum simulations \cite{Nokkala18a},   multi-party quantum communication \cite{Cai17,Arzani19}, and  quantum metrology \cite{Pinel12,Gessner18}.

For all the applications that rely on the transfer of information over appreciable distances, telecommunication wavelengths offer the most reliable solution. In this regime single-mode quantum states with large amount of squeezing  \cite{Ast13,Gehring15}, and on-chip  few-modes squeezed states have been reported \cite{Mondain19,Lenzini18}. Second-order non-linear waveguides are promising candidates for the controlled generation of both single- and  multimode quantum states at telecom wavelengths \cite{Eckstein11,Harder16,Brecht15,Ansari18}, but a direct characterization of CV multimode quantum resources, like cluster states, in these systems is still missing. Waveguides offer in fact large nonlinearities and long interaction lengths, along with  the possibility to tailor the modal decomposition of the quantum state \cite{Dirmeier20}, but losses and modal engineering have to be carefully controlled  in order to measure the desired quantum effects via homodyne detection used in CV protocols.

In this work we study the generation of CV multimode entangled states at  telecom wavelengths via parametric down conversion (PDC) in non-linear waveguides. We aim at obtaining a controlled number of modes with sufficient squeezing, which can coherently interfere in order to generate the wanted entangled states via schemes exploiting both the spectral  \cite{SPOPO} and the temporal \cite{Yokoyama13} degrees of freedom of a femtosecond  laser source \cite{LaVolpe20}. We identify type \emph{II} PDC in periodically poled non-linear waveguides in the so-called symmetric group velocity matching (SGVM) configuration \cite{ChristineLaserPhysics} as the most promising scheme. 

The paper is structured as follows. In Section II, we set the theoretical grounds, define the physical variables and discuss the optimal configuration of the non-linear waveguide. In Section III, we revisit the concepts underlying SGVM and show the properties of type \emph{0}, \emph{I} and \emph{II} PDC processes, concluding that type \emph{II} is best suited for the target application. Section IV applies the results to potassium titanyl phosphate (KTP) and yields realistic fabrication parameters. Section V finishes with the conclusions.

\section{PDC as a continuous variables quantum resource}

\subsection{Squeezed modes}

The parametric interaction in a $\chi^{(2)}$ non-linear crystal involving a pump field and two output fields, usually called signal and idler, can be expressed in appropriate conditions (see for example \cite{Horoshko19}) via an effective hamiltonian $\hat{\mathrm{H}}$ that governs the evolution of the quantum operators in the interaction picture. This effective hamiltonian can be written (assuming a classical undepleted pump), as:
\begin{equation}
        \mathrm{\hat{H}}=\sum_{i,j}J_{i,j}\hat{a}^\dag_i\hat{b}^\dag_j + \mathrm{h.c.},
    \label{H_eff}
\end{equation}
where $J_{i,j}$ are the elements of the so-called joint two-photon amplitude   and $\hat{a}_i^\dag$ and $\hat{b}_j^\dag$ are creation operators of, respectively, the signal and idler photons in spatio-temporal modes $f_i(\mathbf{r},t)$ and $l_j(\mathbf{r},t)$, each forming an orthonormal mode basis.

The discrete form of the hamiltonian in Eq.~(\ref{H_eff}) is due to the implicit discrete mode basis $\{f_i(\mathbf{r},t), l_j(\mathbf{r},t)\}$ in which it is expressed. In this work, we will restrict ourselves to temporal modes, assuming a single spatial mode common for all the fields involved, a configuration that is typically realised in waveguides. Under these conditions we write $f_i(\mathbf{r},t)\rightarrow f_i(t)
\xleftrightarrow{F.T.} \tilde{f}_i(\omega)$, where $\omega$ is the frequency and $F.T.$ stands for Fourier Transform.

In the continuous frequency basis, the effective hamiltonian of Eq.~(\ref{H_eff}) is written as:
\begin{equation}
    \mathrm{\hat{H}}=\int\int{\mathrm{d}\omega_s\mathrm{d}\omega_i J(\omega_s,\omega_i)\hat{a}^\dag(\omega_s)\hat{b}^\dag(\omega_i)} + \mathrm{h.c.},
    \label{H_eff_freq}
\end{equation}
where now the joint two-photon amplitude is a two-dimensional function of the signal and idler frequencies $\omega_s$ and $\omega_i$, known as the joint spectral amplitude \cite{Grice97}, and the creation operators create field excitations at different frequencies.

The joint spectral amplitude, $J(\omega_s,\omega_i)$, contains information about the pump field that enters the waveguide and the linear and non-linear optical properties of the waveguide itself. One can perform a decomposition of this function in a set of modal functions $\{h_k(\omega_s)\}$ and $\{g_k(\omega_i)\}$ \cite{Law00}:
\begin{equation}
    J(\omega_s,\omega_i) = \sum_{k}\lambda_{k} h_{k}(\omega_s)g_{k}(\omega_i).
    \label{Schm_decom}
\end{equation}
This decomposition is also known as the Schmidt decomposition, and the set of complex values $\{\lambda_k\}$ the Schmidt coefficients, fulfilling $\sum_k|\lambda_k|^2=1$.

By inserting Eq.~(\ref{Schm_decom}) into Eq.~(\ref{H_eff_freq}) we obtain:

\begin{equation}
    \hat{\mathrm{H}} = \sum_{k}\lambda_k\hat{A}^\dag_k\hat{B}^\dag_k + \mathrm{h.c}\\
    \label{H_schmidt}
\end{equation}
\begin{equation}
\begin{aligned}    
    \hat{A}_k^\dag &= \int{\mathrm{d\omega_s}h_k(\omega_s)\hat{a}^\dag(\omega_s)}\\
    \hat{B}_k^\dag &=
    \int{\mathrm{d\omega_i}g_k(\omega_i)\hat{b}^\dag(\omega_i)}
    \label{Schmidt_op}
\end{aligned}
\end{equation}

The Hamiltonian of Eq.~(\ref{H_schmidt}) generates two sets of eigenmodes that are EPR entangled with each other (twin modes), where each couple is in the so called two-mode squeezed state \cite{Migdal10,Horoshko19,Christ13}, and which are defined spectrally by the same sets $\{h_k(\omega_s)\}$ and $\{g_k(\omega_i)\}$ of the Schmidt decomposition in Eq.~(\ref{Schm_decom}). These eigenmodes are also called temporal modes \cite{Brecht15}. Eq.~(\ref{Schmidt_op}) is then simply a basis change from the continuous frequency basis to the discrete Schmidt basis.

From the continuous variable picture of the process, we define the normalized quadrature operators associated to the temporal modes as:
\begin{align}
\label{quad}
\begin{split}
 \{\hat{q}_{A,k},\hat{p}_{A,k}\}=  \{\hat{A}_k^{\dag} +  \hat{A}_k, i(\hat{A_k}^{\dag} -\hat{A_k})\}
\\
 \{\hat{q}_{B,k},\hat{p}_{B,k}\}=  \{\hat{B}_k^{\dag} +  \hat{B}_k, i(\hat{B_k}^{\dag} -\hat{B_k})\}
\end{split}
\end{align}
The EPR couples are then characterized by the following relations of quadrature variances $\Delta^2(\hat{q}_{A,k}-\hat{q}_{B,k})=2 e^{-2 \lambda_{k}}, \Delta^2(\hat{q}_v) =2 e^{-2 \lambda_{k}}  $ and $\Delta^2(\hat{p}_{A,k}+\hat{p}_{B,k})=2 e^{-2 \lambda_{k}}, \Delta^2(\hat{p}_v)= 2 e^{-2 \lambda_{k}} $, where $\Delta^2(\hat{q}_v)$ and $\Delta^2(\hat{p}_v)$ are the variances of the quadratures in the case of vacuum states, that here are taken equal to one according to the quadrature definition in Eq.~(\ref{quad}). This means that in the case of high gain, \emph{i.e.}, high value for the Schmidt coefficients $\lambda_{k}$, the quantum state of each EPR couple is an approximation of the so called  quadrature entangled EPR state \cite{EPR}, characterized by  $\hat{q}_{A,k}=\hat{q}_{B,k}$ and $\hat{p}_{A,k}=-\hat{p}_{B,k}$.

It can be shown \cite{Christ11} that the effective number of modes, $K$, \emph{i.e.}, the effective number of pairs of functions from $\{h_k(\omega_s),g_k(\omega_i)\}$, can be calculated as
\begin{equation}
    K = \frac{1}{\sum_k|\lambda_k|^4}.
    \label{K}
\end{equation}
This quantity is often called the Schmidt number, and it will be a figure of merit in our analysis.

The temporal modes form eigenbases of the signal and idler parts of the total field. One can show \cite{Horoshko19} that the eigenvalues of the total joint spectral amplitude are doubly degenerate, with eigenvectors $\{s_k(\omega)\}$ such that:
\begin{equation}
    s_k(\omega) = \frac{1}{\sqrt{2}}(h_k(\omega) \pm g_k(\omega)),
    \label{squeezing_modes}
\end{equation}
which are mutually uncorrelated modes that present squeezing. These modes are called squeezed modes (or ``supermodes'' \cite{Patera10}), and their creation operators are defined as $\hat{S}^{\dag}_{k \pm}=\frac{1}{\sqrt{2}}(\hat{A}^{\dag}_k \pm \hat{B}^{\dag}_k$). The effective hamiltonian in terms of the supermode creation operators is then:
\begin{equation}
    \hat{\mathrm{H}} = \sum_{k}\frac{\lambda_k}{2}\left(\left(\hat{S}^\dag_{k,+}\right)^2+\left(\hat{S}^\dag_{k-}\right)^2\right) + \mathrm{h.c}\\
    \label{H_supermodes}
\end{equation}
The corresponding variance of the quadrature variables, with operators  $ \{\hat{q}_{S,k},\hat{p_{S,k}}\}$, defined in the same way as in Eq.~(\ref{quad}), are $\Delta^2(\hat{q}_{S,k})= e^{-2 \lambda_{k}}$ and  $\Delta^2(\hat{p}_{S,k})= e^{2 \lambda_{k}}$. They are squeezed, and their level of squeezing, $2 \lambda_{k}$, is proportional to the corresponding Schmidt eigenvalue $\lambda_k$.

Hence, in the temporal mode basis, the signal and idler parts of the total field are individually diagonalized and there exists entanglement between them, while in the squeezed mode basis, the modes are uncorrelated and carry squeezing as a quantum resource.

It is important to note that the transformation of Eq.~(\ref{squeezing_modes}) physically corresponds to the interference of the signal and idler modes on a balanced beamsplitter.

In the following, we focus on the calculation and characterization of the squeezed eigenmodes $s_k(\omega)$ and their squeezing eigenvalues $\lambda_k$, since they constitute a natural multimode quantum state where continuous variable quantum information can be encoded and processed. In particular, the Bloch-Messiah reduction \cite{BlochMessiah} ensures that all multimode Gaussian states, \emph{i.e.}, the states whose quadratures are characterized by Gaussian statistics and that can be generated by hamiltonians that are at most quadratic in the bosonic operators like the ones in Eqs. (\ref{H_eff}), (\ref{H_eff_freq}) (\ref{H_schmidt}) and (\ref{H_supermodes}), can be written as a set of single mode squeezed states and two multiport interferometers. In our context, this means that we could generate deterministically any highly entangled multimode Gaussian state if we input the squeezed eigenmodes into an appropriate multiport interferometer, generating a configurable source of cluster states that could then be used as the resource for measurement based quantum computation, multiparty quantum communication or quantum metrology.

\subsection{Use of waveguides and mode detection}

Until now we have considered the general case of non-degenerate parametric down conversion and the emergence of the squeezed modes. We will now specify and justify the configuration that will be treated in the rest of the paper.

We will consider the use of non-linear waveguides, instead of bulk crystals, for their ability to confine light over a long length (in particular over the coherence length of the pump laser) and their discretization in the number of available spatial modes. Both effects  enhance the non-linear interaction by several orders of magnitude. This enables the generation of single-pass squeezing \cite{Dirmeier20}; no optical cavities are required. The use of waveguides also means that we will work in the so-called collinear configuration, where signal and idler fields exit the crystal in the same beam. 

Another important reason for the use of waveguides is that, as we will see later, the engineering of the physical parameters of the waveguide opens up new possibilities for engineering the PDC, and hence the generated modes, which will benefit our target applications. 

Furthermore, the measurement of the modes would be performed via coherent detection, in particular homodyne detection, where the quadratures of the targeted field can be accessed by interfering the field with a local oscillator. Hence, in order to measure the quadratures of the different squeezed modes, the local oscillator should be shaped spectrally to match the corresponding modes $\{s_k(\omega)\}$. However, this becomes a difficult task if the squeezed modes take arbitrary shapes. In Eq.~(\ref{squeezing_modes}) we mentioned that the squeezed modes are obtained by the interference of signal and idler temporal modes, which can themselves be approximated as Hermite Gauss modes \cite{Ansari18}. In order for this interference to be efficient and for the squeezed modes to inherit the approximate Hermite Gauss functional form, both signal and idler temporal modes should be as similar as possible, which translates in having indistinguishable fields after the non-linear interaction in the waveguide.

The spatial indistinguishability is given due to the collinear configuration imposed by the use of waveguides, while in the bulk crystal scenario a precise alignment of the crystal would be needed. The spectral indistinguishability constrains to work in the so-called degenerate case, where both central frequencies for signal and idler are equal (and equal  to half the pump central frequency by conservation of energy): $\omega_s^{0} = \omega_i^{0} = \omega_p^{0}/2$.

The final degree of freedom of the fields is their polarization. Depending on the output polarization in the PDC process, three types of parametric interactions can be found. In type \emph{0} and type \emph{I}, the polarization of signal and idler fields is the same, while in type \emph{II} the fields have orthogonal polarizations, making them distinguishable. However, we will show in the next section that, due to their dispersion properties, both type \emph{0} and type \emph{I} prove unpractical. We will consequently restrict ourselves to type \emph{II} processes. We note that the polarization distinguishability can be erased by rotating the polarizations of the signal and idler by 45 degrees and interfering them on a polarizing beamsplitter. At the output of the beamsplitter, one then obtains the squeezed modes \cite{Dirmeier20}. We also note here that, due to the different polarizations of the signal and idler modes, there may be a temporal walk-off after the waveguide, which can be corrected for with appropriate compensation crystals.

\begin{figure}
\centering
\includegraphics[width=0.5\textwidth]{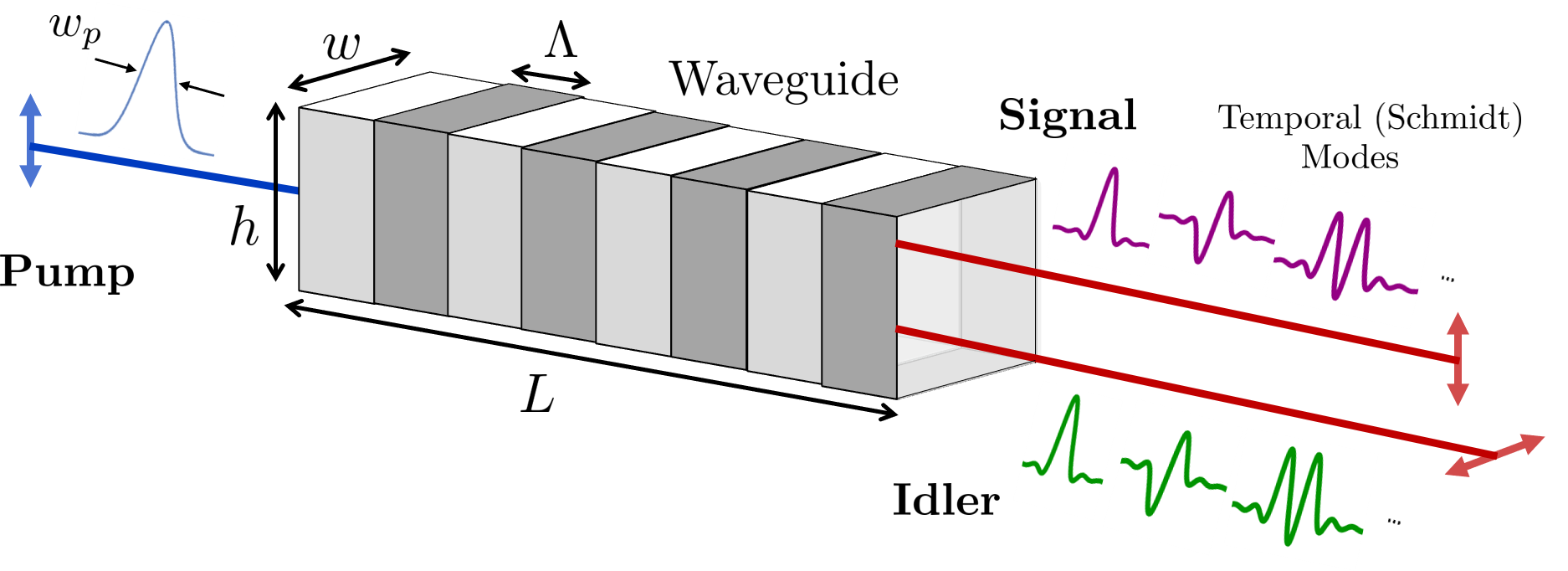}
\caption{Periodically poled non-linear waveguide
scheme picturing degenerate, collinear, type \emph{II} parametric downconversion. The squeezed modes are obtained after making the two sets of temporal (Schmidt) modes interfere. For more details see text.}
\label{waveguide_scheme}
\end{figure}

Fig.~\ref{waveguide_scheme} schematically shows the configuration discussed above: degenerate collinear type \emph{II} parametric down conversion in a non-linear waveguide.

\section{Symmetric group-velocity matching}

\subsection{Joint spectral amplitude structure }

The mathematical structure of the joint spectral amplitude, or JSA, of Eq.~(\ref{H_eff_freq}), $J(\omega_s,\omega_i)$, can be expressed as:
\begin{equation}
J(\omega_s,\omega_i) = \alpha_p(\omega_s+\omega_i)\phi(\omega_s,\omega_i),
\label{jsa}
\end{equation}
where the subscripts $s/i/p$ stand for signal, idler, and pump, and $\omega_p = \omega_s+\omega_i$. The function $\alpha_p(\omega_s+\omega_i)$ is the pump envelope function in the spectral domain, which we take to be a Gaussian with a certain width, denoted by $w_p$ (see Fig.~\ref{waveguide_scheme}). Finally, $\phi(\omega_s,\omega_i)$ is the phasematching function. It is governed by the waveguide's optical properties and the geometry of the problem. This function does not generally have properties of symmetry.

The phasematching function can be written as:
\begin{equation}
\phi(\omega_s,\omega_i) = \mathrm{sinc}\left(\frac{L}{2}\Delta k(\omega_s,\omega_i)\right),
\label{phasematching_function}
\end{equation}
where $L$ is the waveguide length and the function $\Delta k(\omega_s,\omega_i)$ is the wavevector mismatch
\begin{equation}
    \Delta k(\omega_i, \omega_s) = k_p(\omega_i+\omega_s) - k_s(\omega_s) - k_i(\omega_i),
    \label{mismatch}
\end{equation}
where the $k$ are the corresponding wavevectors. In general, Eq.~(\ref{mismatch}) is a vectorial equation, but we recover its scalar version because of the collinear condition.

It is possible to match two arbitrary signal and idler frequencies so that $\Delta k=0$ with a well-known technique known as quasi-phasematching \cite{quasiphasematching}, which introduces a new term to Eq.~(\ref{mismatch}) by modulating the $\chi^{(2)}$ coefficient along the waveguide with a poling period $\Lambda$. Eq.~(\ref{mismatch}) including the quasi-phasematching modulation can be written as:

\begin{equation}
    \Delta k(\omega_i, \omega_s) = k_p(\omega_i+\omega_s) - k_s(\omega_s) - k_i(\omega_i) - \frac{2\pi}{\Lambda},
    \label{mismatch_qpm}
\end{equation}
such that the selection of the poling period makes the mismatch zero for a certain pair of desired frequencies. In this work, the poling period will correspond to the case of matched signal/idler central frequencies.

\subsection{Type \emph{0} and \emph{I} PDC}

Indistinguishability of the signal and idler fields is required for their proper interference after the non-linear interaction. This implies that the joint spectral amplitude has to be symmetric under the exchange of the signal and idler labels, \emph{i.e.},
\begin{equation}
    J(\omega_s,\omega_i) = J(\omega_i,\omega_s).
    \label{jsa_sym}
\end{equation}
Since the pump function $\alpha_p(\omega_i + \omega_s)$ is naturally symmetric under this exchange, the fields are indistinguishable if the phasematching function shares this symmetry, \emph{i.e.}, if $\phi(\omega_s,\omega_i) = \phi(\omega_i,\omega_s)$. From Eq.~(\ref{phasematching_function}) we therefore require the condition on the wavevector mismatch: $\Delta k(\omega_s,\omega_i) = \Delta k(\omega_i,\omega_s)$. 

In the case of type \emph{0} and type \emph{I} PDC, it is easy to see that the symmetry condition is fulfilled, since $k_s(\omega_s) = k_i(\omega_i)$ in these cases.

As in \cite{ChristineLaserPhysics}, we write the Taylor series of the mismatch function around the central frequencies $\omega_s^{0}= \omega_i^{0}=\omega_p^{0}/2$ to second order:
\begin{equation}
\begin{aligned}
    \Delta k =& \Delta k^0 + \gamma_s(\omega_s-\omega^0_p/2) + \gamma_i(\omega_i-\omega^0_p/2) +\\ & \delta_s(\omega_s-\omega^0_p/2)^2 + \delta_i(\omega_i-\omega^0_p/2)^2 +\\ &
    \delta_p(\omega_s-\omega^0_p/2)(\omega_i-\omega^0_p/2)+...
    \label{taylor_exp}
\end{aligned}
\end{equation}
The coefficients are related to the derivatives of the wavevector with respect to the frequency as:
\begin{equation}
\begin{aligned}
   \gamma_{s/i} &= \frac{\partial k_p}{\partial \omega}\Bigr|_{\substack{\omega_p}} - \frac{\partial k_{s/i}}{\partial \omega}\Bigr|_{\substack{\omega_p/2}}\\
   \delta_{s/i} &= \frac{\partial^2 k_p}{\partial \omega^2}\Bigr|_{\substack{\omega_p}} - \frac{\partial^2 k_{s/i}}{\partial \omega^2}\Bigr|_{\substack{\omega_p/2}}\\
   \delta_p &= 2\frac{\partial^2 k_p}{\partial \omega^2}\Bigr|_{\substack{\omega_p}},
    \label{taylor_coeff}
\end{aligned}
\end{equation}
where we wrote $\omega_p$ instead of $\omega_p^{0}$ in the evaluation of the derivatives for clarity in the notation. The constant term $\Delta k^0 = \Delta k(\omega_s^0,\omega_i^0)$ can be made zero via quasi-phasematching, as described previously. 

The coefficient $\gamma_s$ ($\gamma_i$) associated with the linear term in the expansion is related to the pump and signal (idler) group velocities, while the coefficients $\delta_s$ ($\delta_i$) associated with the quadratic term is related to the pump and signal (idler) group velocity dispersion.

In type \emph{0} and type \emph{I} PDC, $\gamma_i=\gamma_s\equiv \gamma$, $\delta_i = \delta_s\equiv \delta$, etc. In addition, usually $w_p\leq10\,$~nm, such that the linear term dominates. Hence, we can approximately write:

\begin{equation}
    \Delta k(\mathrm{type\ \emph{0}/\emph{I}}) = \gamma(\omega_s+\omega_i-\omega_p^0)+ \mathcal{O}(\omega_s^2,\omega_i^2,\omega_s\omega_i)
    \label{mismatch_type0I}
\end{equation}
The phasematching can be approximated by a function of $\omega_s+\omega_i$, and therefore so can the joint spectral amplitude $J(\omega_s+\omega_i) = \alpha_p(\omega_s+\omega_i)\phi(\omega_s+\omega_i)$. 

\begin{figure}
\centering
\includegraphics[width=0.5\textwidth]{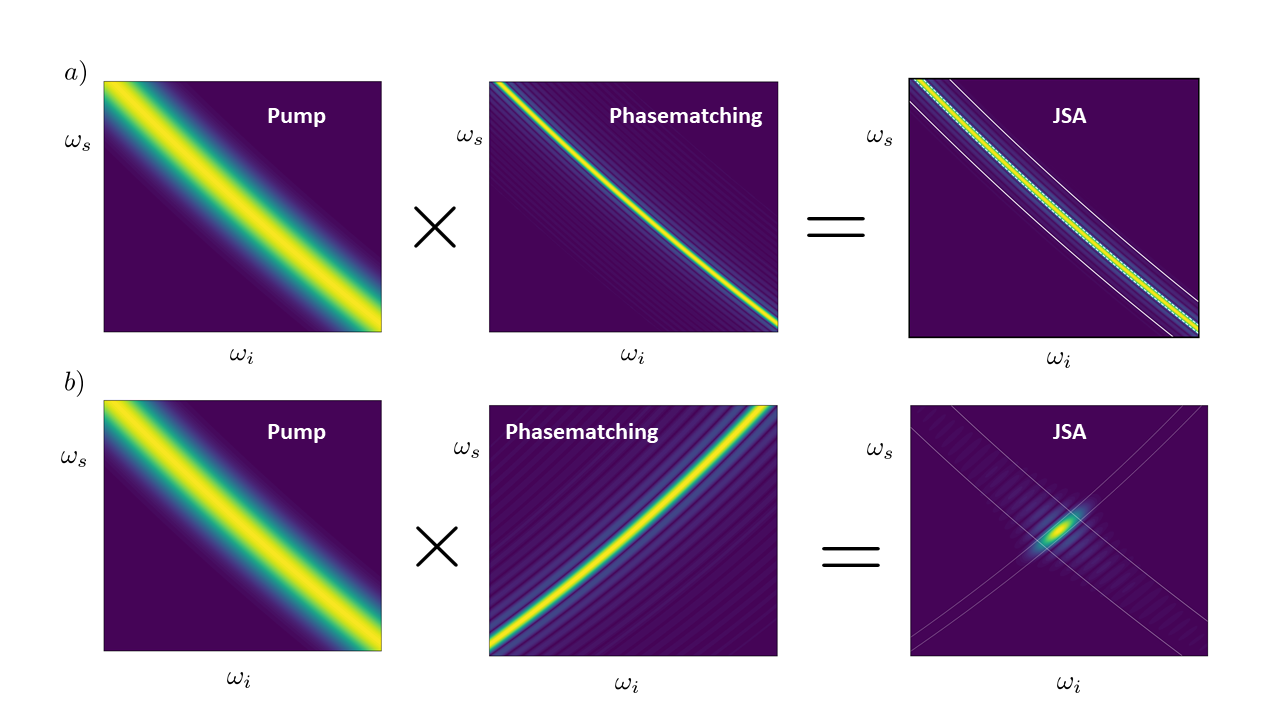}
\caption{Pump, phasematching and JSA functions in degenerate collinear PDC, a) type \emph{0}/\emph{I}, b) type \emph{II}.}
\label{JSAs}
\end{figure}

Since both the pump envelope and the phasematching are functions of the sum of the frequencies, they are aligned, making a $-$45$^\circ$ angle, in the 2-dimensional coordinate system with axes ($\omega_s$,$\omega_i$). The JSA is then non zero in a large spectral region, yielding spectrally broad temporal modes that are not well-suited for homodyne detection, since one needs good spectral overlap with a local oscillator. Fig.~\ref{JSAs} a) shows this situation. Overcoming this problem necessitates either very broad local oscillators and/or higher order term contribution in the mismatch function, both of which is unpractical.

\subsection{Type \emph{II} PDC}

For type \emph{II} PDC, the signal and idler polarizations are orthogonal, the fields are distinguishable and hence their dispersion relations are different, $k_s(\omega_s)\neq k_i(\omega_i)$. 

However, looking at Eq.~(\ref{taylor_exp}), we obtain identical modes if we manage to have the condition $\gamma_s = -\gamma_i$, which is called the symmetric group velocity matching (SGVM) condition \cite{ChristineLaserPhysics}. This translates into having the following condition on the group velocities (recalling that the group velocity is $v_g(\omega) = \partial \omega/\partial k$):
\begin{equation}
    \frac{1}{v_{g,p}(\omega_p^0)} = \frac{1}{2}\left(\frac{1}{v_{g,s}(\omega_s^0)} + \frac{1}{v_{g,i}(\omega_i^0)}\right)
\end{equation}

In that case, the Taylor expansion of Eq.~(\ref{taylor_exp}) gives:
\begin{equation}
    \Delta k(\mathrm{type\ \emph{II}}) = \gamma(\omega_s-\omega_i)+ \mathcal{O}(\omega_s^2,\omega_i^2,\omega_s\omega_i),
    \label{mismatch_typeII}
\end{equation}
with $\gamma \equiv \gamma_s = -\gamma_i$. Hence, the mismatch function is antisymmetric under the exchange of signal and idler labels: $\Delta k(\omega_s,\omega_i) = -\Delta k(\omega_i,\omega_s)$. By Eq.~(\ref{phasematching_function}) and the even parity of the sinc function we obtain a symmetric phasematching function, and hence a symmetric joint spectral amplitude (up to first order in the Taylor series of Eq.~(\ref{taylor_exp})). Under this condition, signal and idler fields are indistinguishable (except for their polarization).

From Eq.~(\ref{mismatch_typeII}), the joint spectral amplitude is approximately of the form
\begin{equation}
    J(\omega_i,\omega_s) = \alpha_p(\omega_i+\omega_s)\phi(\omega_s-\omega_i).
\end{equation}
In Fig.~\ref{JSAs} b) we see that the phasematching function now makes an angle of $+45^\circ$ in the ($\omega_s$,$\omega_i$) coordinate system, hence making a $90^\circ$-angle with the pump envelope function. Therefore, the JSA is non zero only in a small spectral region around the central frequencies. In this situation, the coherent detection of the squeezed modes is experimentally realizable and the interference of signal and idler fields is approximately perfect after the change of polarization of one of the two.

In conclusion, all the reasoning exposed until now leads us to consider the configuration of degenerate, collinear, type \emph{II} PDC in non-linear waveguides. We will also consider being around the symmetric group velocity matching condition, which means:

\begin{itemize}
    \item The poling period $\Lambda$ provides quasi-phasematching for the central frequencies $\omega^0_p$ and $\omega_s^0 = \omega_i^0 = \omega_p^0/2$.
    \item The inverse group velocities at the central frequencies satisfy the condition $\gamma_s = -\gamma_i$, where these coefficients are defined in Eq.~(\ref{taylor_coeff}).
    \item The higher order terms in the Taylor expansion of Eq.~(\ref{taylor_exp}) are sufficiently small compared to the first order term in the entire frequency range considered (see \ref{hig-ord}  for insights into this point).
\end{itemize}
We will also call the wavelength fulfilling these conditions as symmetric group velocity matching wavelength $\lambda_{\mathrm{SGVM}}$.

\subsection{Index of refraction}
The dispersion relation
\begin{equation}
    k(\omega) = \frac{\omega}{c}n(\omega),
    \label{wav_index}
\end{equation}
where $c$ is the speed of light in vacuum and $n$ the index of refraction, encapsulates the optical properties of the crystal and could depend on several physical variables apart from frequency, such as temperature, $T$, and waveguide characteristics. The latter include the waveguide section size, with height $h$ and width $w$, and spatial mode order inside the waveguide, defined by two integer numbers $n_1$ and $n_2$.

If one is interested in modeling the functional form of the index of refraction with these waveguide parameters, approximations must be made. A wide-spread approximation is the so-called metallic waveguide approximation \cite{Benniethesis}, in which one assumes the waveguide to be surrounded by perfectly conducting edges. This is the approximation that will be made in this work. More refined modeling of the index of refraction involves more advanced computational methods solving Maxwell equations inside and outside the waveguide, like Mercatili's method~\cite{Marcatili} or finite element methods~\cite{FiniteElement}. 

Using the metallic waveguide approximation, the index of refraction is written as
\begin{equation}
    n(\lambda, T, n_1, n_2, w, h)=n(\lambda,T) + \left(\lambda\frac{n_1+1}{2h}\right)^2 + \left(\lambda\frac{n_2+1}{2w}\right)^2,
    \label{index_Ref}
\end{equation}
where the function $n(\lambda,T)$ depends on the non-linear material under consideration, and is given by the empirical Sellmeier equations (for e.g. KTP see \cite{Takaoka}).

In this work, we will restrict ourselves to the case in which only the fundamental spatial mode travels through the waveguide, which translates into $n_1=n_2=0$. This is exactly the case in a perfect single-mode waveguide and, for shorter wavelengths, this can be achieved by careful mode-matching in the input coupling.

Furthermore, we will have a different index of refraction of the form of Eq.~(\ref{index_Ref}) for different light polarizations. For instance, in the case of uniaxial crystals, we would have two indices of refraction (for the ordinary and extraordinary axis), and in biaxial crystals, like KTP, we would have one index of refraction for every spatial direction. 

These indices define the different wavevectors for pump, signal and idler fields in each case. We can therefore study the symmetric group velocity matching condition via the dependence of the indices of refraction on wavelength, temperature and waveguide dimensions.

\subsection{Mode overlap}
To wrap up this section, we define the quantity that will quantify the similarity between signal and idler temporal modes as the overlap integral $o_n$.
 The $n$-th order overlap between two modal functions, $h_n(\omega)$ and $g_n(\omega)$, is defined as:
\begin{equation}
\begin{aligned}
    o_n &= \frac{1}{N}\Big\lvert\int_{-\infty}^{\infty}\mathrm{d}\omega h^*_n(\omega)g_n(\omega)\Big\rvert\\
    N &= \sqrt{\int_{-\infty}^{\infty}\mathrm{d}\omega|h_n|^2(\omega)\int_{-\infty}^{\infty}{d}\mathrm{d}\omega|g_n|^2(\omega)},
    \label{overlap}
    \end{aligned}
\end{equation}
where $N$ is a normalization factor, such that the overlap is adimensional and its maximum value is 1, if and only if $h_n(\omega)=g_n(\omega)$. 

\section{Results}
In our analysis, we will focus on KTP as the non-linear material for the waveguide. Lithium niobate (LN) is another important non-linear material, since commercial LN waveguides already exist. However, calculations with LN in our configuration lead to the conclusion that the waveguide width and height should be in the order of 1 $\mu$m for the symmetric group velocity condition to hold. This is smaller than the telecom wavelengths considered here (around 1.55 $\mu$m). The metallic waveguide approximation breaks down at this scale and more elaborate techniques must be used in order to model the index of refraction of these small structures. Therefore, we will only consider KTP in this work, which naturally provides SGVM at telecom wavelenghts.

\subsection{SGVM wavelength}

First, we performed calculations to compute the wavelengths fulfilling the SGVM condition. This first section shows the existence of this wavelength for KTP for reasonable physical parameters.

For this, we fixed the temperature and the waveguide dimensions to typical values and computed the dispersion properties of the crystal, in particular the different group velocities and group dispersions in KTP, from the indices of refraction of Eq.~(\ref{index_Ref}), as a function of wavelength. Thanks to the metallic waveguide approximation, the derivatives can be computed analytically in this case. In this way the coefficients of Eq.~(\ref{taylor_coeff}), as well as the corresponding poling period, are calculated as a function of wavelength.

\begin{figure}
\centering
\includegraphics[width=0.5\textwidth]{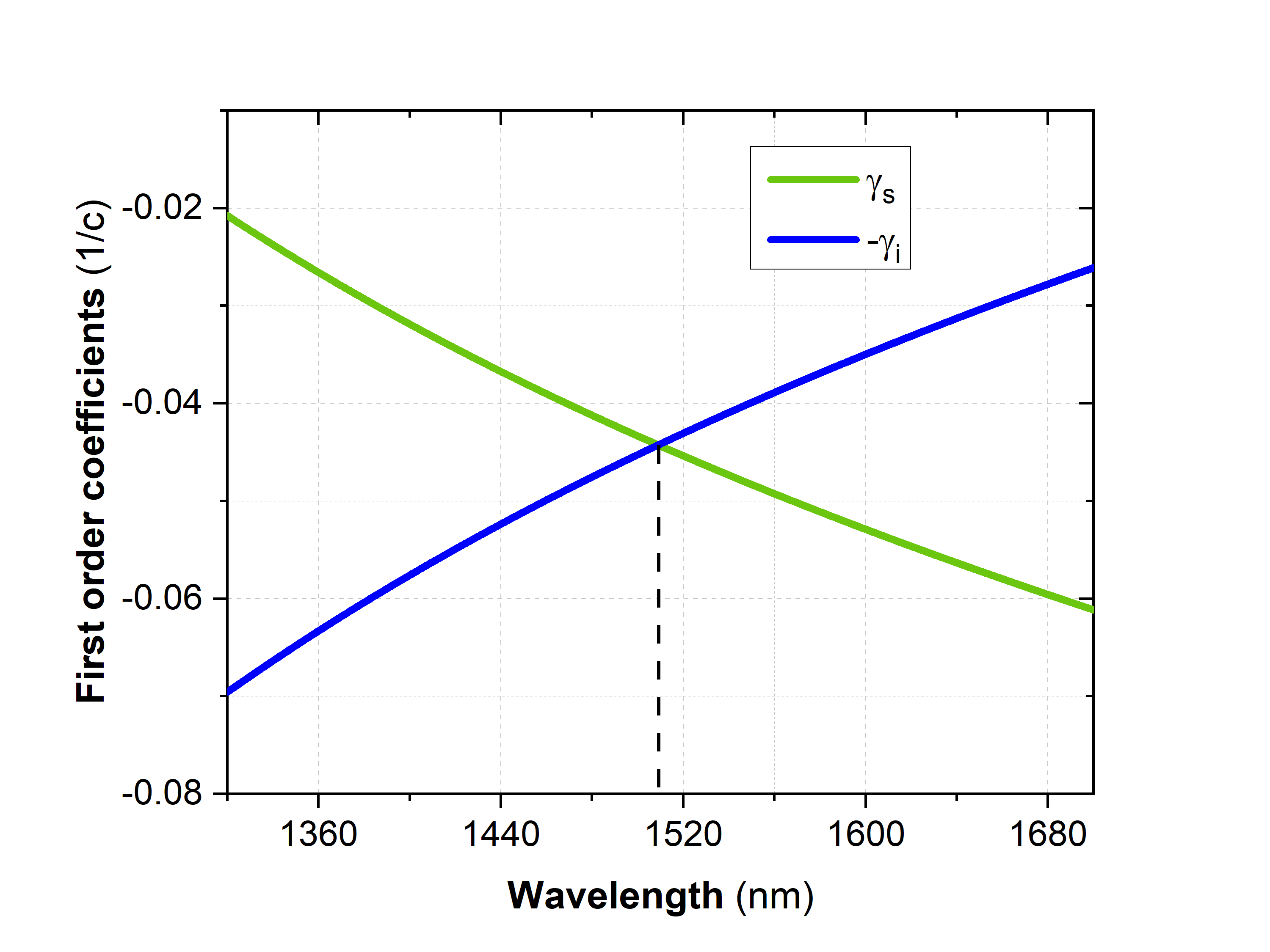}
\caption{Taylor coefficients for the linear term of Eq.(\ref{taylor_exp}), defined in Eq.(\ref{taylor_coeff}). The two curves cross at a certain wavelength where the SGVM condition $\gamma_s=-\gamma_i$ is fulfilled.(KTP at $T=20\ ^o\mathrm{C}$ and $w,h=6\ \mathrm{\mu m}$)}
\label{First_order_coeff}
\end{figure}

Fig.~\ref{First_order_coeff} shows the Taylor coefficients associated with the group velocities ($\gamma_s$ and $-\gamma_i$). They coincide at the SGVM phasematching wavelength, $\lambda_{\mathrm{SGVM}}$, where $\gamma_s=-\gamma_i$.

Once $\lambda_{\mathrm{SGVM}}$ is obtained, it is important to check the contribution of the higher order terms in Eq.~(\ref{taylor_exp}). This is discussed in \ref{hig-ord}.

\subsection{Telecom wavelength}

In the calculations above, a wavelength fulfilling the SGVM condition was proven to exist for KTP, Fig.~\ref{First_order_coeff}. We will now be interested in varying the other physical variables affecting the index of refraction in Eq.~(\ref{index_Ref}), namely temperature and waveguide dimensions, to get the condition at a certain desired wavelength. 

In particular, we are interested in obtaining SGVM at 1550 nm, given that telecom wavelengths present low losses in optical fibers, and there exist optimized of-the-shelf components at this wavelength, opening the way to practical applications.

It is important to remark the geometrical choice of the problem, because results would be different depending on the orientation of the crystal's optical axes with respect to the pump field polarization. In particular, the Kleinman symmetry \cite{Boyd} provides the supported processes for all  combinations in propagation and polarization directions. The only configuration in which we have results at wavelengths around 1550 nm is the one that corresponds to a crystal cut (\emph{i.e.}, a vertical polarization direction) in the $z$ direction and propagation direction along the $x$ direction, where the indices of refraction for every spatial direction are ordered like in \cite{Takaoka}. Therefore this is the case considered in this work. More details can be found in~\ref{Klein}. 

As announced above, we will now consider the dependence of the index of refraction on temperature and waveguide dimensions.

\begin{figure}
\centering
\includegraphics[width=0.5\textwidth]{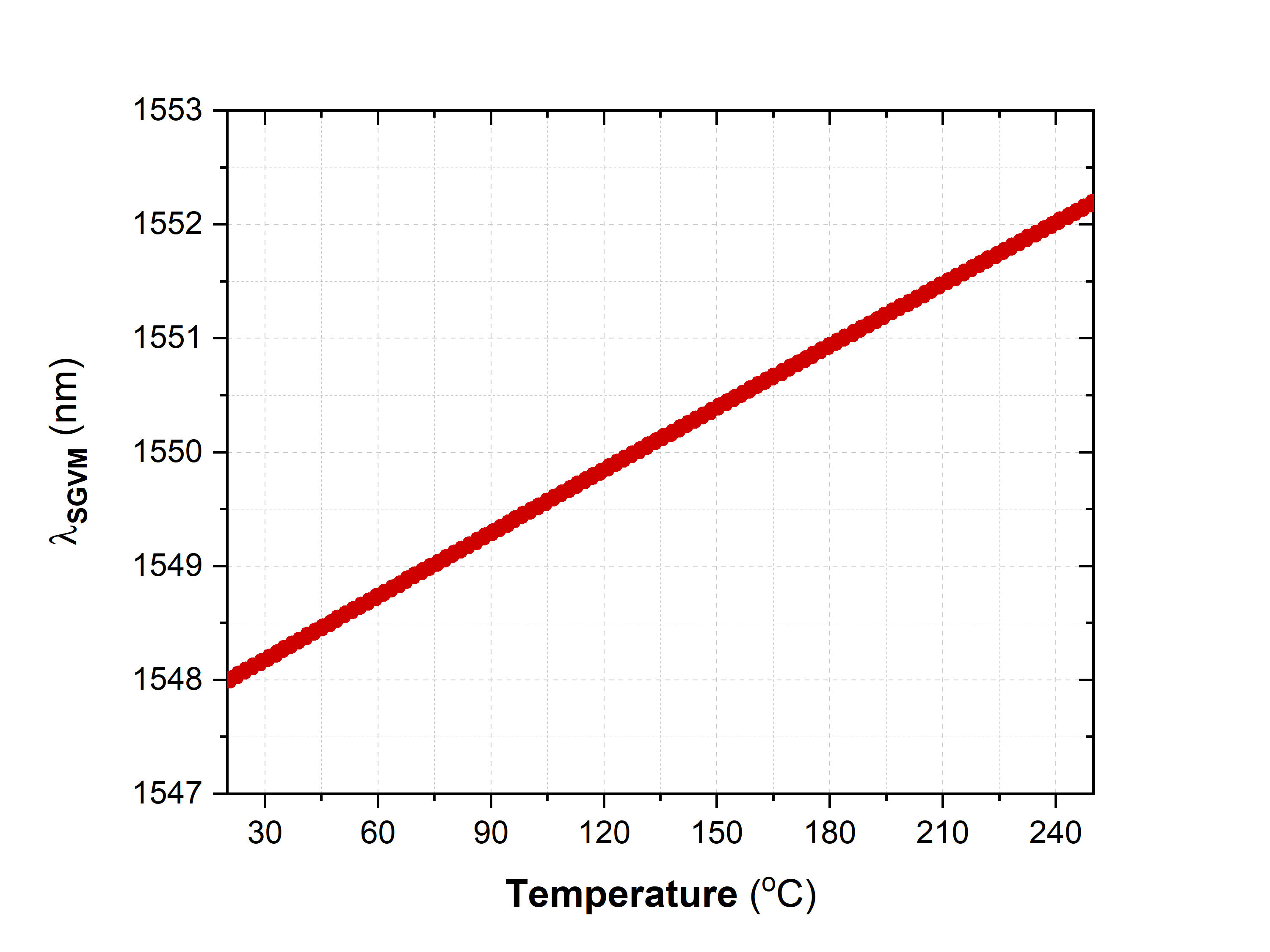}
\caption{Temperature dependence of the SGVM wavelength for KTP. The waveguide dimensions were set to $w=h=9\ \mathrm{\mu m}$. }
\label{temperature_lambdaep_ktp}
\end{figure}

\paragraph{Temperature:} Fig.~\ref{temperature_lambdaep_ktp}
shows the temperature dependence of the SGVM wavelength for KTP. For this calculation we fixed the waveguide dimensions such that $\lambda_{\mathrm{SGVM}}\simeq$ 1550 nm.

We obtain a change in the SGVM wavelength of $\Delta\lambda_{SGVM} \simeq 4$ nm in a range of temperatures of $\Delta T \simeq 230$ $^o$C.  This result indicates that temperature has a very limited effect on the change of $\lambda_{\mathrm{SGVM}}$. 

According to these results, it seems reasonable to fix the temperature and focus on the tuning of the waveguide dimensions. We therefore fixed the temperature to be room temperature for the rest of the work, which is natural when considering future technology applications, where involved temperature control may be undesirable.

\begin{figure}
\centering
\includegraphics[width=0.55\textwidth]{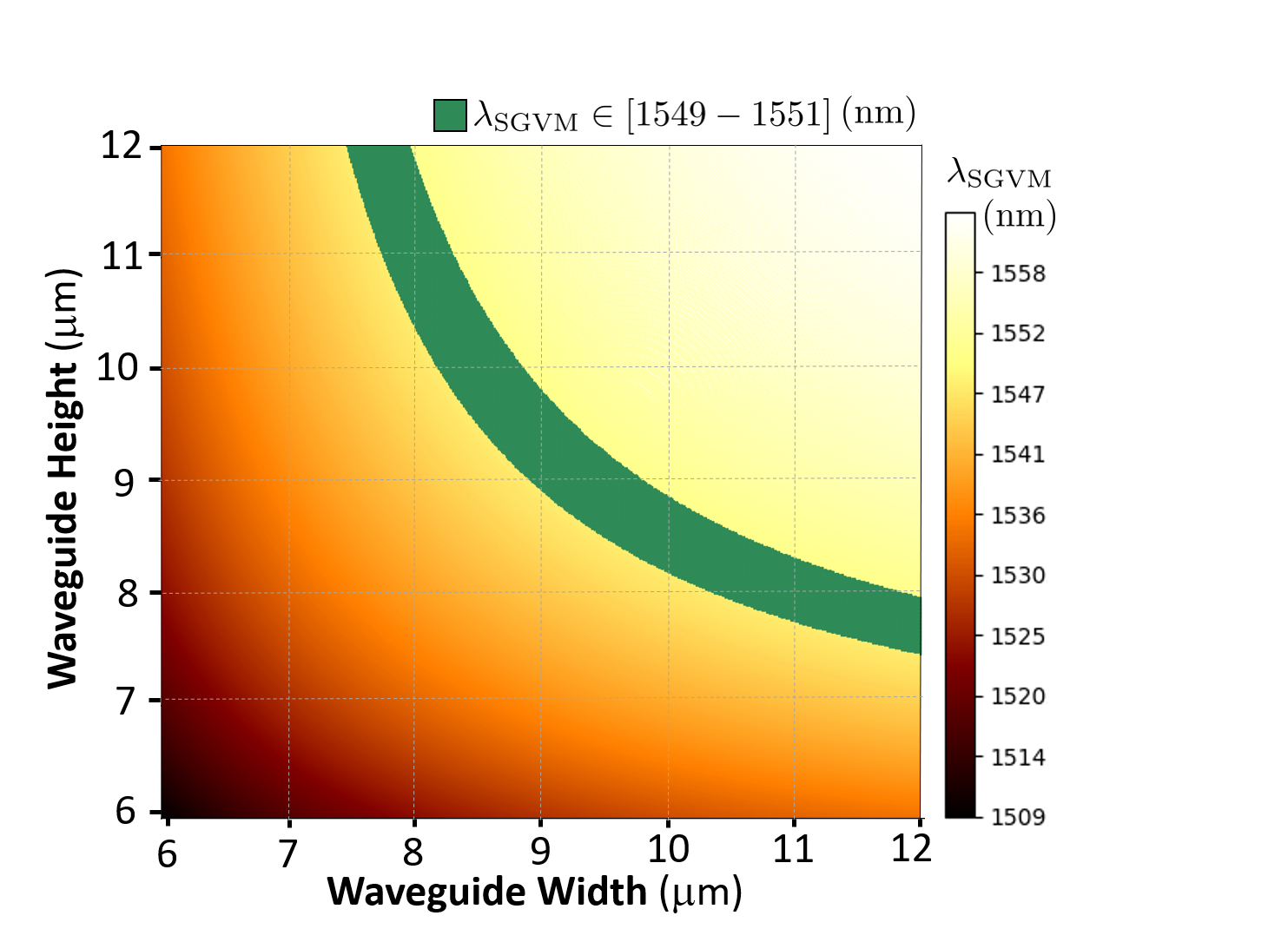}
\caption{SGVM wavelength as a function of waveguide height and width for KTP. The green area indicates the region where $\lambda_{\mathrm{SGVM}}\in[1549-1551]$ nm. The temperature was set to $T=20\ ^o\mathrm{C}$.}
\label{lambdaep_wgdim_ktp}
\end{figure}

\paragraph{Waveguide Dimensions:} Fig.~\ref{lambdaep_wgdim_ktp} shows the SGVM wavelength as a function of the waveguide width and height for KTP. As the waveguide size increases, the surface flattens, making apparent the diminishing impact of the waveguide dispersion. We also find this behaviour in Eq.~(\ref{index_Ref}), where the waveguide dimensions contribution to the index of refraction disappears as $h,w\rightarrow\infty$. The green area in Fig.~\ref{lambdaep_wgdim_ktp} marks values of $\lambda_{\mathrm{SGVM}}$ that lie in the interval between 1549 and 1551 nm. Therefore, we conclude that to work at 1550 nm in KTP, the waveguide should have a size of about 9 by 9 $\mu$m in width and height. From Fig.~\ref{temperature_lambdaep_ktp} we know that this is almost independent of temperature in the temperature interval considered.

\subsection{Engineering of the quantum states}

Next, we performed simulations of the PDC process for KTP to compute the signal and idler modes of our system with the associated squeezing eigenvalues under the SGVM condition. We remark that this exact procedure could be done for any non-linear optical material if its dispersion properties are known.

Our pump function is modeled as a (square-normalized) Gaussian function in the following way: 
\begin{equation}
    \alpha_p(\omega_s + \omega_i) = \frac{1}{\sqrt{\sqrt{\pi}w_p}}\exp\left(-\frac{(\omega_p^0-(\omega_s+\omega_i))^2}{2w_p^2}\right),
\end{equation}
where the parameter $w_p$ is the pump width. The pump central wavelength is $\lambda_p^0 = 775$ nm, corresponding to a central frequency $\omega_p^{0}=2\pi c/\lambda_p^{0}$. Furthermore, in our results, we will express the pump width in length units (wavelength bandwidth instead of frequency bandwidth), by redefining $w_p$ as $2\pi c/w_p$.

The phasematching function is that of Eq.~(\ref{phasematching_function}), where we fix the waveguide length. The wavevector mismatch is calculated using Eqs.~(\ref{wav_index}) and (\ref{index_Ref}), where we also set values for the temperature and waveguide dimensions in each case. The poling period is calculated for every case so that the signal/idler central wavelengths, $\lambda_s^0 = \lambda_i^0 = 2\lambda_p^0 = 1550$ nm, are perfectly phasematched.

With these two functions we compute the joint spectral amplitude, or JSA, according to Eq.~(\ref{jsa}). 

We then perform the Schmidt decomposition of Eq.~(\ref{Schm_decom}) by computing the singular value decomposition of the JSA, obtaining two matrices containing the signal and idler temporal modes and a diagonal matrix with the Schmidt coefficients $\{\lambda_k\}$. We recall that the distribution of the squeezing eigenvalues is equivalent (except for a global factor of 2) to the distribution of the Schmidt coefficients. From these coefficients we obtain the Schmidt number according to Eq.~(\ref{K}) and we also compute the temporal modes' full width at half maximum (FWHM). This yields the spectral width of the temporal modes, which is important for designing their coherent detection. Finally, we use Eq.~(\ref{overlap}) to obtain the overlap between the signal and idler modes in order to quantify their similarity. 

The main goal of this analysis is to obtain the dependence of the quantum fields on input parameters that will be specified in the following. For the purpose of this work, we can see the computation just described as a set of input variables that are experimentally controllable, and output variables after the parametric interaction. The main input variables are the waveguide length, the waveguide dimensions (height and width), and the pump width, and the main output variables are the Schmidt number, the modes FWHM and their overlap. We then perform distinct simulations to obtain the corresponding dependencies. Note that the temperature could be taken as another input variable. However, following the discussion earlier, we consider it to be a fixed variable (at room temperature), since we have seen that it would have a minor effect on the output modes.

For the sake of clarity, we construct Table \ref{Tabla}, where each column represents an input variable  and each row a corresponding output variable. The table elements show general dependencies that will be discussed in detail in the following. 

\begin{table}
\caption{Columns are input variables and rows are outputs. For each element, the description expresses the output/input dependency in the PDC process.\\}
\begin{tabular}{ |c|c|c|c| } 
\hline
 & \multirow{2}{5em}{\textbf{Pump Width}} &\multirow{2}{7em}{\textbf{Waveguide Dimensions}} & \multirow{2}{5em}{\textbf{Waveguide Length}} \\
  & & & \\
\hline
\multirow{2}{5em}{\textbf{Schmidt Number}} & \multirow{2}{7.5em}{{Minimum and linear growth}} & $\sim$ Constant & Grows \\ 
& & & \\ 
\hline
\multirow{2}{5em}{\textbf{Mode FWHM}} & Grows & Inverse for & Decreases  \\ 
& & signal/idler & \\ 
\hline
\multirow{2}{5em}{\textbf{Overlap}} & Constant & SGVM & Constant  \\ 
& & Analysis & \\ 
\hline
\end{tabular}
\label{Tabla}
\end{table}

The range of values adopted for the input variables is the following: we consider pump widths from 1 to 12 nm, waveguide dimensions between 4 and 10 $\mathrm{\mu m}$, and waveguide lengths between 5 and 30 mm. The pump width can be modified by tuning the pump field that enters the waveguide, while the waveguide dimensions and length are fixed during the waveguide fabrication process.

\begin{figure}
\centering
\includegraphics[width=0.5\textwidth]{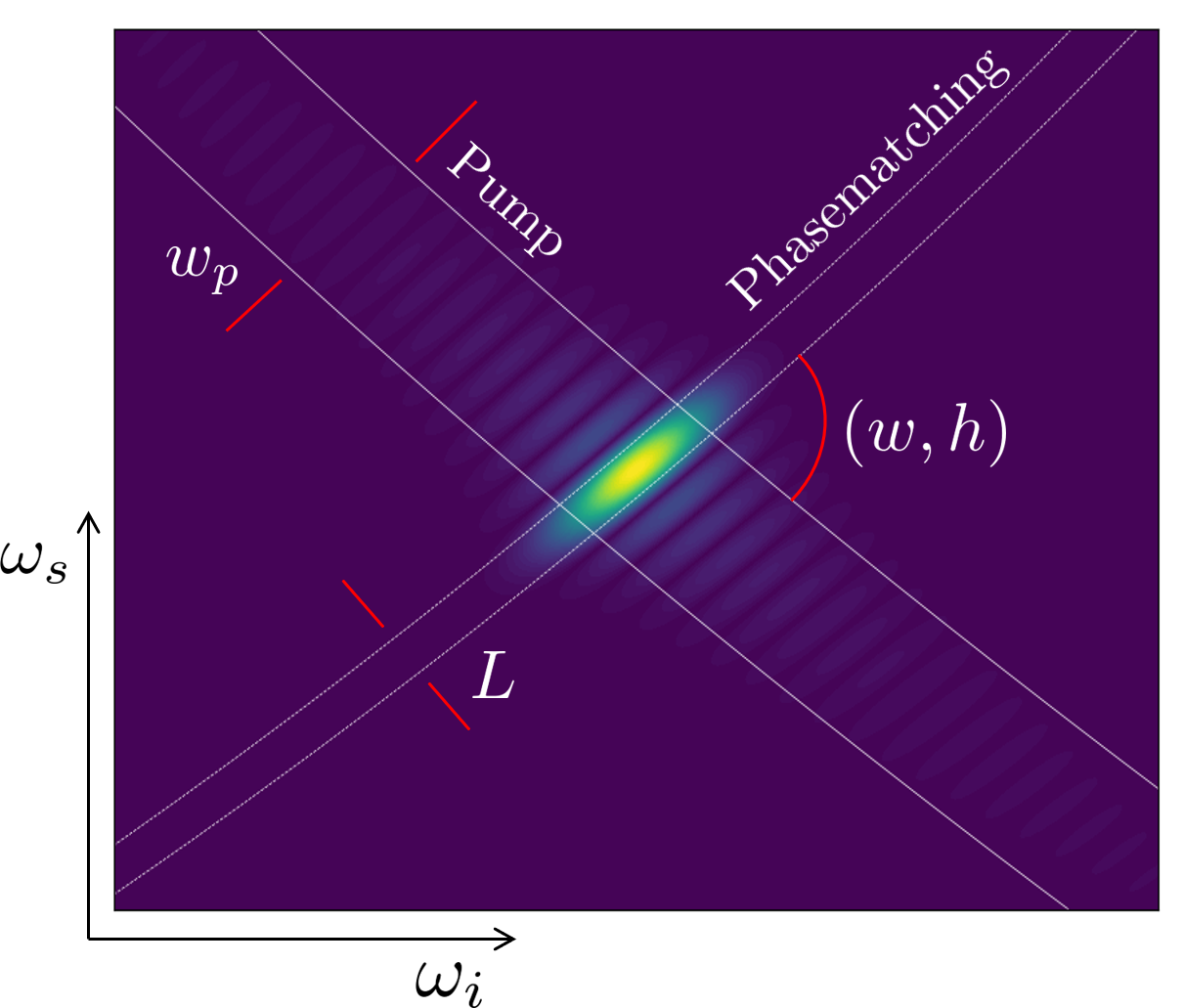}
\caption{Example of a JSA function, highlighting the impact of the waveguide length $L$, pump width, $w_p$, and waveguide dimensions $(w,h)$. For more details see text.}
\label{JSAsum}
\end{figure}

The rest of this section will explore Table \ref{Tabla} in detail. As a general comment, the dependencies are based on the following observations:
\begin{itemize}
    \item The waveguide length, $L$, changes the bandwidth of the phasematching function as $\mathrm{sinc}\left({L^{-1}}\right)$.\\
    \item The width of the pump, $w_p$, although self-explicative, changes the width of the pump function.\\
    \item The waveguide dimensions mainly change the angle between the pump function and the phasematching function, because they change the optical properties (index of refraction) of the material. This is why they are the critical parameter in order to obtain the SGVM condition and why it is interesting considering them as an engineering parameter for the quantum states. We have also seen in this work that even though temperature also takes part in the index of refraction, its contribution is negligible in the usual experimental range of values. We remark that given fixed waveguide dimensions and temperature, the angle between the phasematching function and the pump function is totally defined by the choice of the non-linear material.\\
\end{itemize}
The results of this section can be understood keeping in mind these three observations, which are schematically shown in Fig.~\ref{JSAsum}.

For the sake of clarity, we will refer to the elements of Table \ref{Tabla} like matrix elements. For instance, the element 2.1 refers to the second row and first column, which corresponds to the dependence of the mode FWHM on the pump width.

\paragraph*{First row of Table \ref{Tabla}, the Schmidt number:}

\begin{figure}
\centering
\includegraphics[width=0.5\textwidth]{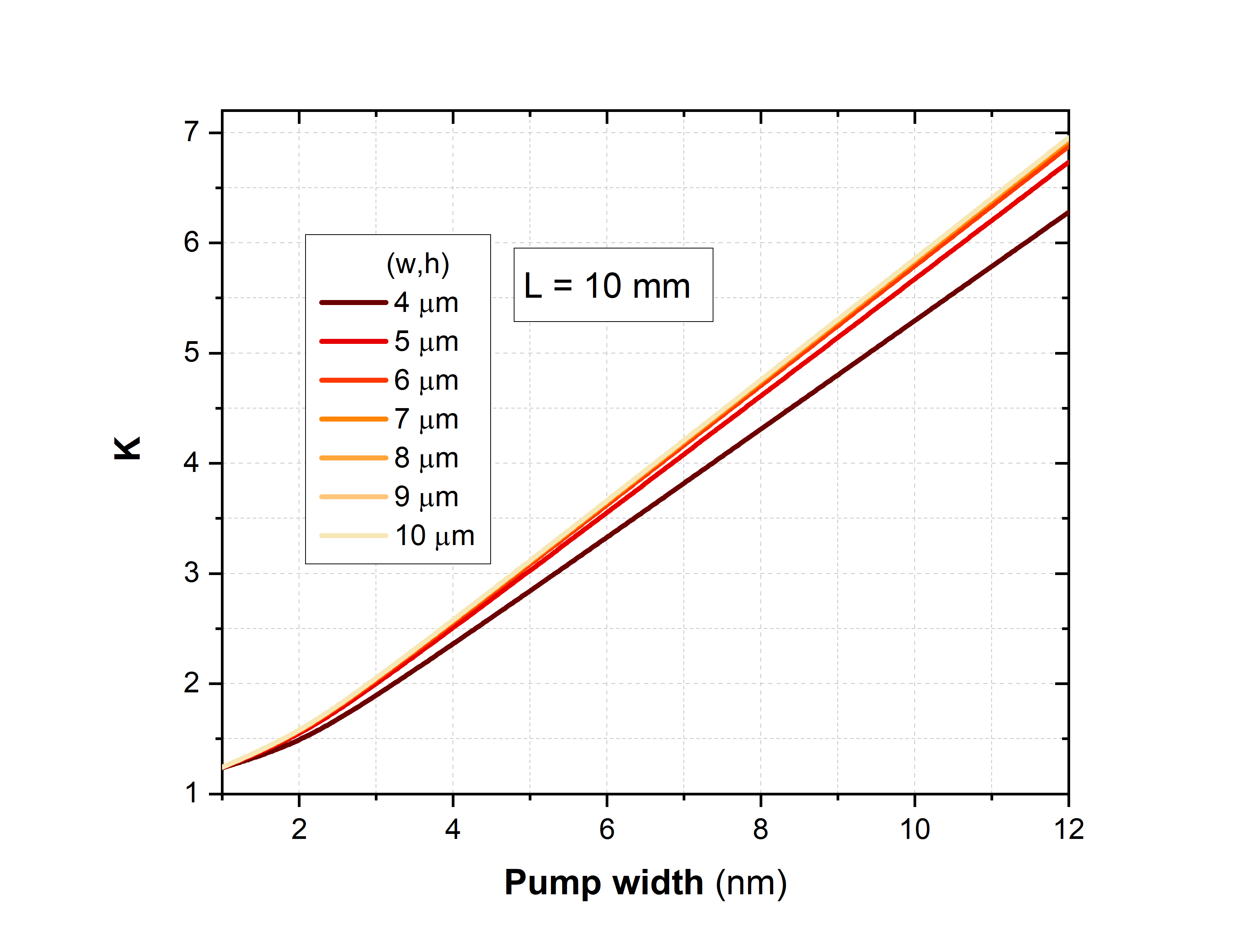}
\caption{Schmidt number as a function of pump width for different waveguide dimensions. Corresponding to the elements 1.1 and 1.2 of Table \ref{Tabla}.}
\label{K_wp}
\end{figure}

Fig.~\ref{K_wp} shows the Schmidt number as a function of pump width for different waveguide dimensions (elements 1.1 and 1.2 of Table \ref{Tabla}). The Schmidt number, and thus the number of modes, grows linearly with the pump width, while the waveguide dimensions have almost no influence. The waveguide length was set to $L=10$ mm for this simulation. This result can be interpreted in the following way: as the pump width increases, the pump envelope does too, and hence its intersection with the phasematching envelope. This creates frequency correlations between signal and idler fields and hence increases $K$. This is true because in this set-up the pump envelope is bigger at any time than the phasematching envelope. In general, there would be a minimum value of $K$ when both envelopes have equal widths.

\begin{figure}
\centering
\includegraphics[width=0.5\textwidth]{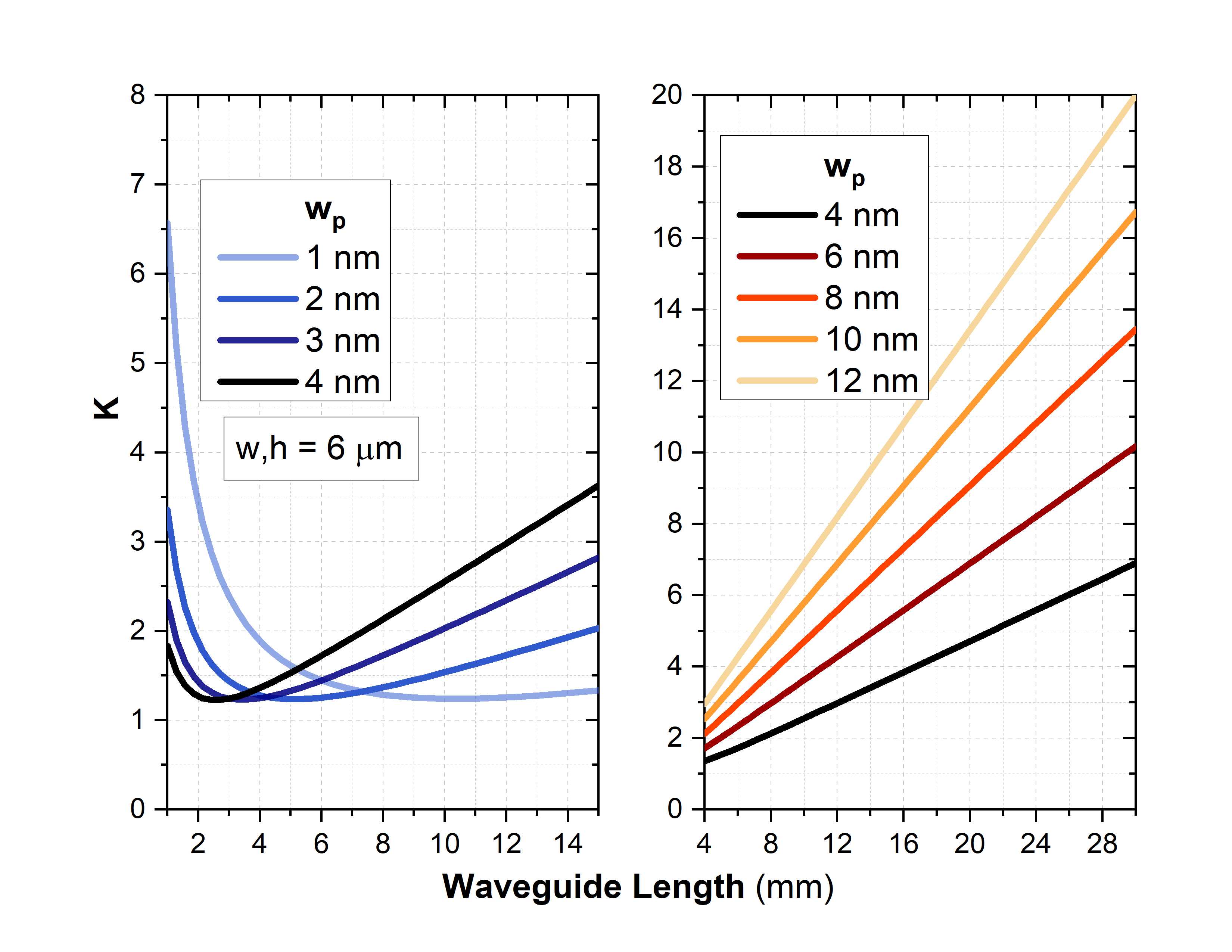}
\caption{Schmidt number as a function of waveguide length for different pump functions. The waveguide dimensions were set to $w,h=6\ \mathrm{\mu m}$. Element 1.3 from Table \ref{Tabla}.}
\label{K_L}
\end{figure}

Fig.~\ref{K_L} shows the Schmidt number as a function of the waveguide length for different pump widths. In this case we are changing the phasematching envelope, and we observe the minimum value of $K$ when both the pump envelope's and the phasematching envelope's widths are equal. This is why we see that the waveguide length in which the minimum value is reached depends on the pump width. Also, the minimum value itself would approximately equal 1 when the intersection of $\alpha_p$ and $\phi$ was at $90^\circ$ that is, at the SGVM condition. This is interesting for applications in single photon sources, where $K = 1$ ensures maximum purity of heralded single photons.

After the minimum point, increasing the waveguide length decreases the width of the phasematching envelope, and so we are effectively in the same situation as in Fig.~\ref{K_wp}, and we observe the same type of behaviour, which is a linear increase in $K$. It is important to note that in Fig.~\ref{K_L} we see that one can control the slope of the increasing linear regime by controlling the pump width, which opens the possibility of modifying the number of modes just by adjusting the pump field. In a realistic scenario, the maximum waveguide length would be limited by the optical losses within the waveguide and the size of the available substrate material. State of the art KTP waveguides have losses of 0.25 dB/cm, which would point towards  waveguide lengths in between 10 and 20 mm. This is in accordance with the maximum size of KTP substrates of roughly $30\times30\,$mm$^2$.

Equivalently, the Schmidt number increases rapidly by decreasing the waveguide length before the minimum, because in that case the 1/L dependence on the phasematching width is more important, creating strong wiggles in the sinc function in Eq.~(\ref{phasematching_function}), and in turn producing strong correlations between the output fields, hence increasing $K$.

\paragraph*{Second row of Table \ref{Tabla}, the modes bandwidth:}

We remark that, given the nature of the Hermite Gauss modes, the mode FWHM of interest is only the first mode (that is approximately Gaussian), as the rest would grow in width as $\sqrt{n}$, where $n$ is the mode order. Therefore knowing the first FWHM is sufficient to approximately characterize all of them.

\begin{figure}
\centering
\includegraphics[width=0.5\textwidth]{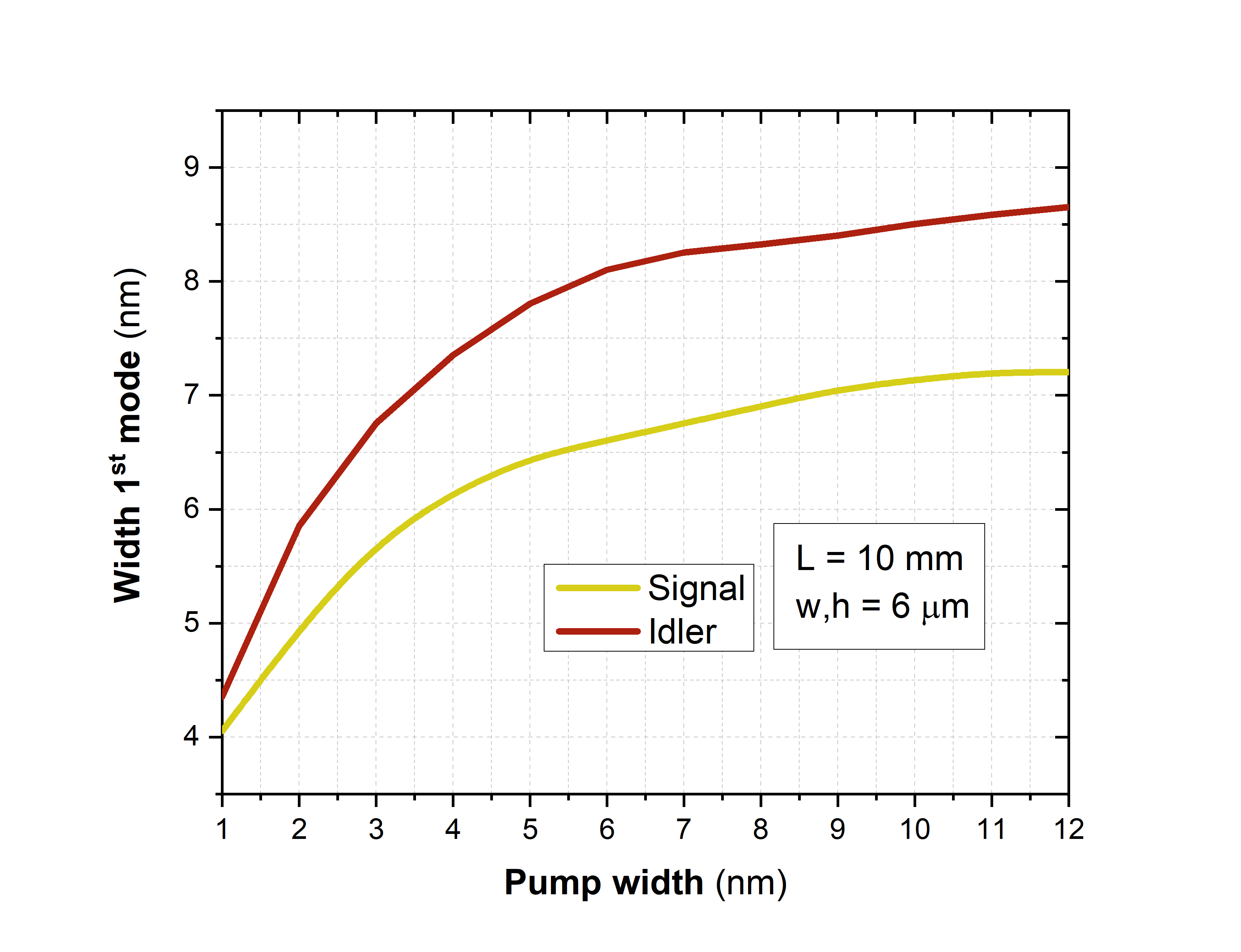}
\caption{First mode FWHM as a function of pump width for a waveguide of $w,h=6\ \mathrm{\mu m}$. Element 2.1 in Table \ref{Tabla}.}
\label{FWHM_wp}
\end{figure}

Fig.~\ref{FWHM_wp} shows the width of this mode as a function of pump width, which corresponds to the element 2.1 in Table \ref{Tabla}. The waveguide dimensions were set to $w,h=6\ \mathrm{\mu m}$, intentionally away from the SGVM condition. We can see that indeed signal and idler modes have different widths for all pump widths and hence their interference would not be perfect with these waveguide dimensions, as expected from the results of the SGVM analysis.

\begin{figure}
\centering
\includegraphics[width=0.5\textwidth]{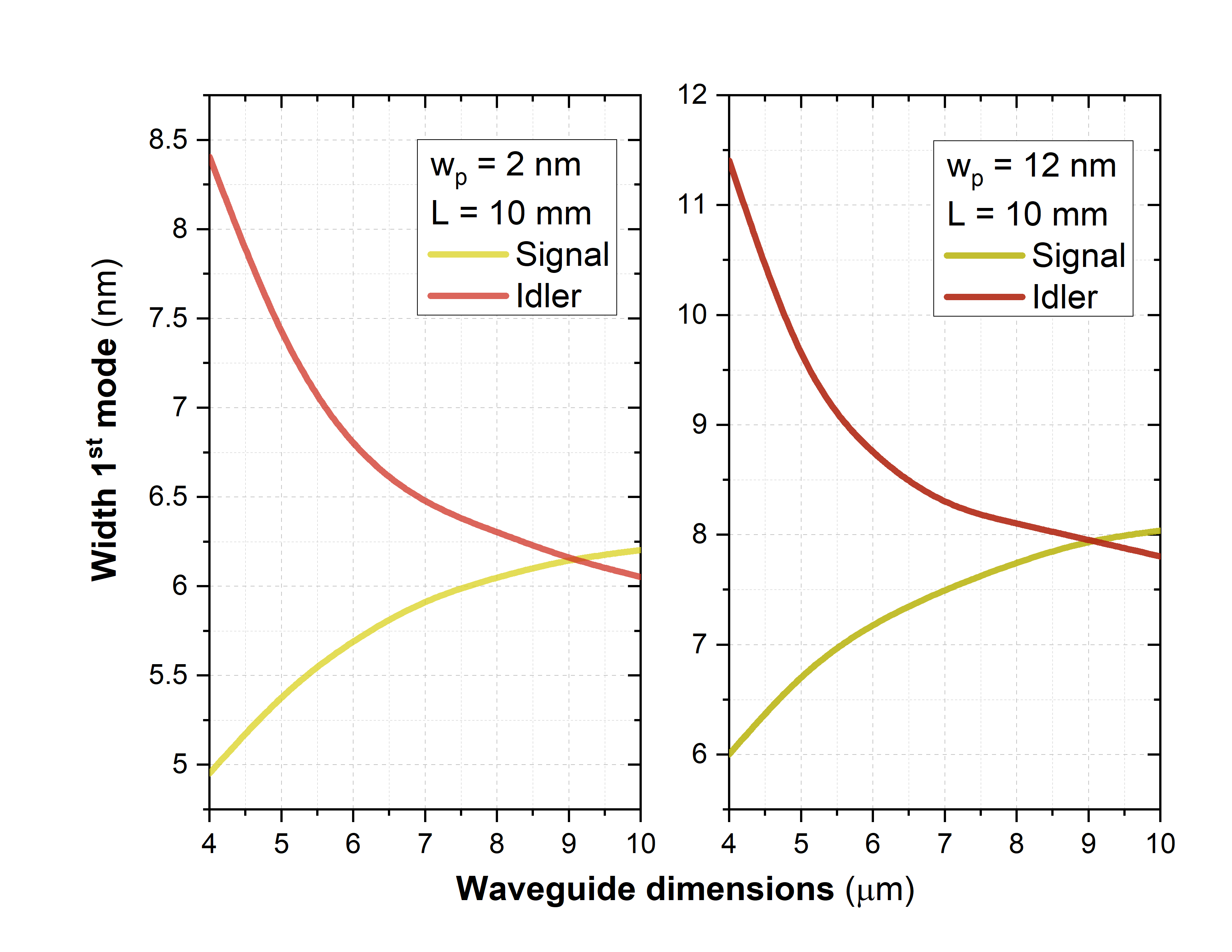}
\caption{First mode FWHM as a function of waveguide dimensions for pump widths of 2 and 12 nm. Elements 2.2 in Table \ref{Tabla}.}
\label{FWHM_wh}
\end{figure}

Fig.\ref{FWHM_wh} shows the modes' FWHM as a function of the waveguide dimensions for pump widths of 2 and 12 nm. Both signal and idler modes coincide at waveguide dimensions near 9 $\mu$m (we considered square waveguides), as expected from the SGVM analysis before. We show the plots for two different pump widths to remark again that the similarity of signal and idler modes is independent of the pump width, as it is related to the waveguide's phasematching function. These relations correspond to the element 2.2 in Table \ref{Tabla}.

\begin{figure}
\centering
\includegraphics[width=0.5\textwidth]{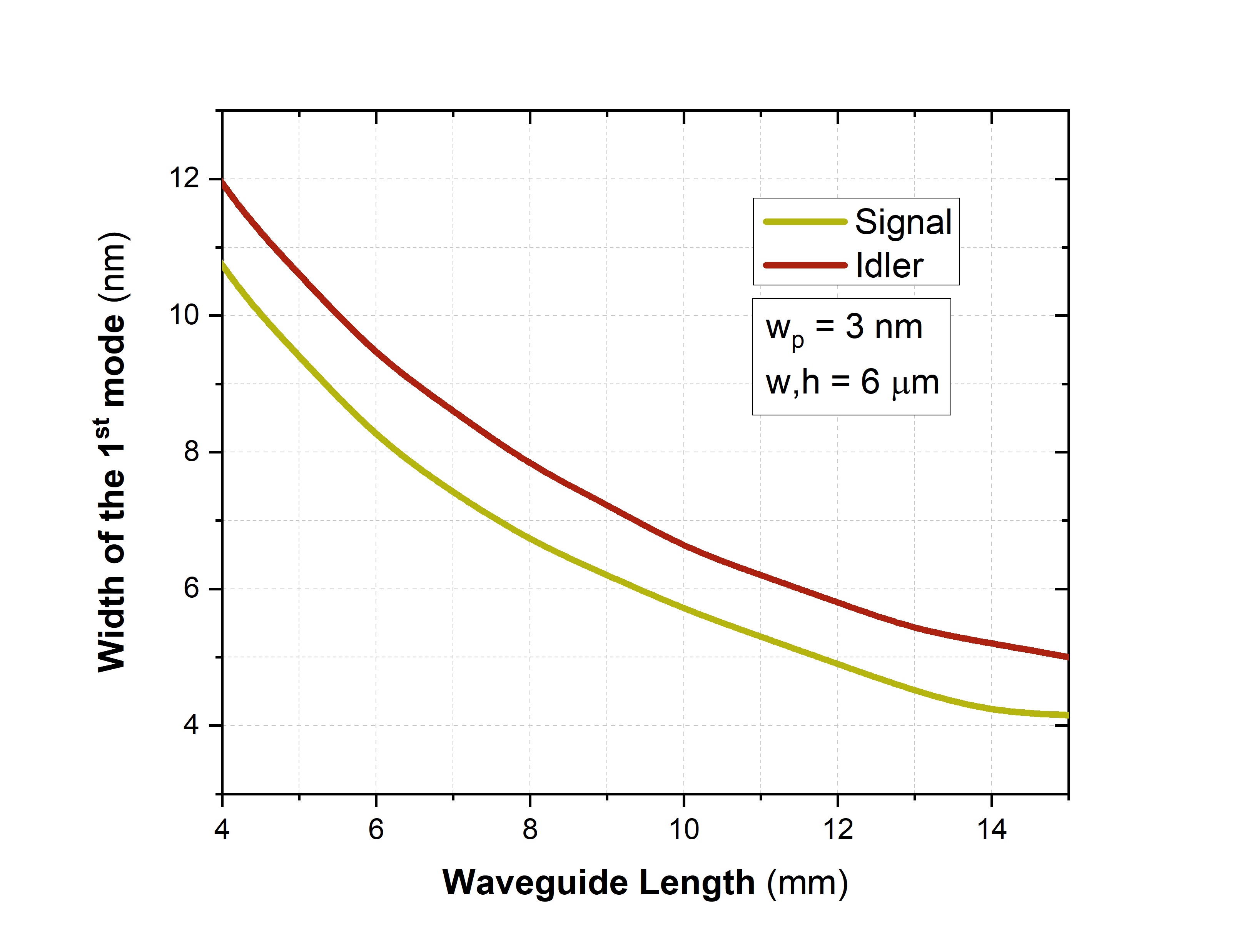}
\caption{First mode FWHM as a function of crystal length with $w_p=3$ nm and $w,h=6\ \mathrm{\mu m}$. Element 2.3 in Table \ref{Tabla}.}
\label{fwhm_L}
\end{figure}

Finally, Fig. \ref{fwhm_L} shows the modes FWHM  as a function of waveguide length for $w_p=3$ nm and $w,h=6\ \mathrm{\mu m}$, again intentionally away from the SGVM condition. Both widths decrease with length, which would point towards the use of the longest possible waveguide if we are to measure the fields by coherent detection. This relation corresponds to element 2.3 in Table \ref{Tabla}.

\paragraph*{Third row of Table \ref{Tabla}, the modes overlap:}

\begin{figure}[H]
\centering
\includegraphics[width=0.5\textwidth]{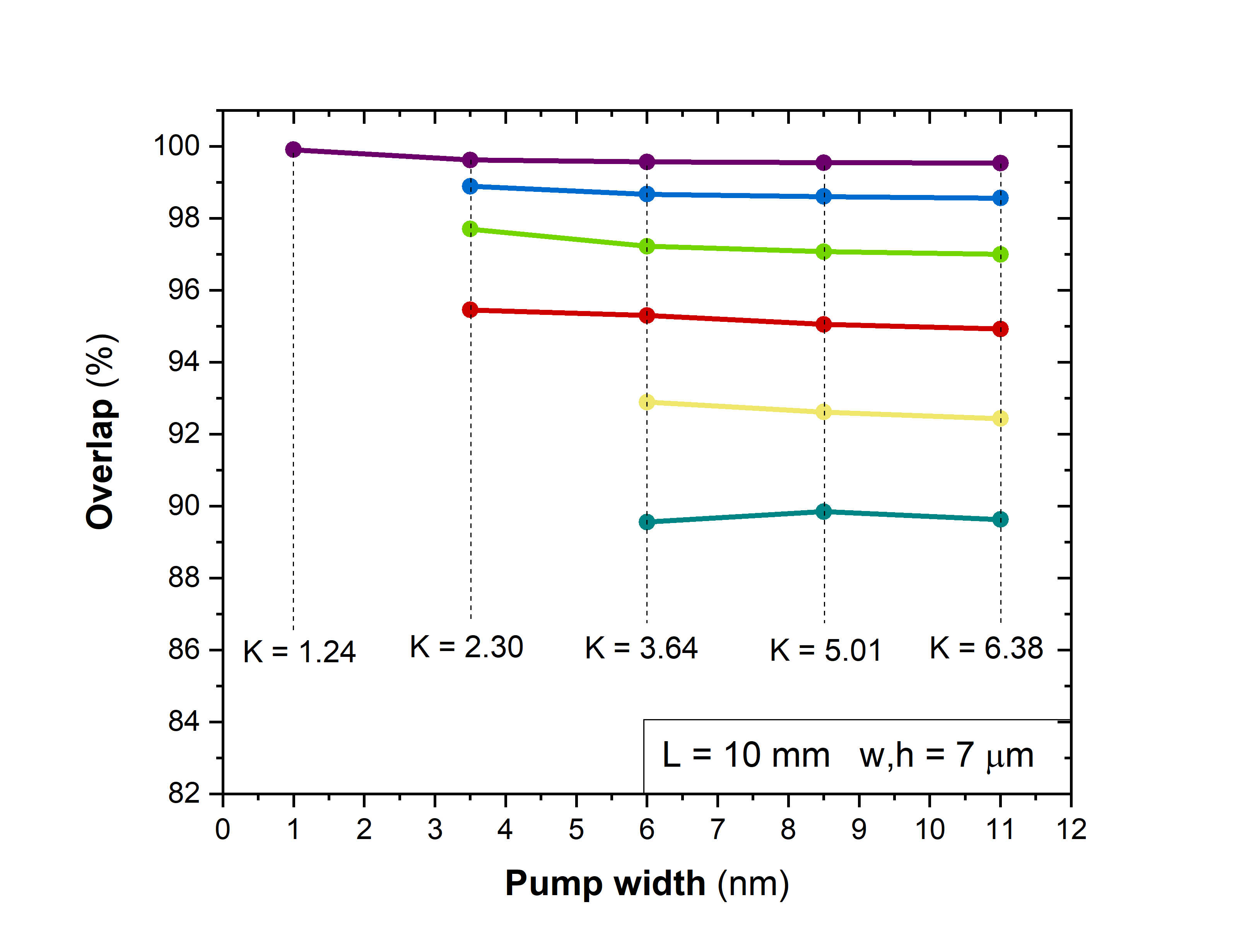}
\caption{Overlap of the first 6 Schmidt temporal modes as a function of pump width. Element 3.1 in Table \ref{Tabla}.}
\label{overlap_wp}
\end{figure}

Fig.~\ref{overlap_wp} shows the overlap of the first six temporal modes as a function of pump width, leaving fixed the rest of the variables. We can see that the overlap decreases with increasing mode order; the reason is discussed in Appendix C, in which we showed that this effect is due to two factors: the increasing width of the modes with the mode order, and the mathematical properties of the Hermite Gauss modes themselves. The pump width does not change the overlap, even though we have just seen that it changes the relative modes' FWHM by up to 25\%. This counter-intuitive behaviour arises due to the stability of the relative width of two Gaussian functions when their relative widths change. The mathematical result is that for a relative width difference of 25\%, \emph{i.e.}, if the width of one Gaussian is 1.25 times the width of the other, then the change in their overlap does not even reach 1\%. This explains why we obtain a constant overlap for different pump widths, although the modes' widths change significantly.

\begin{figure}[H]
\centering
\includegraphics[width=0.5\textwidth]{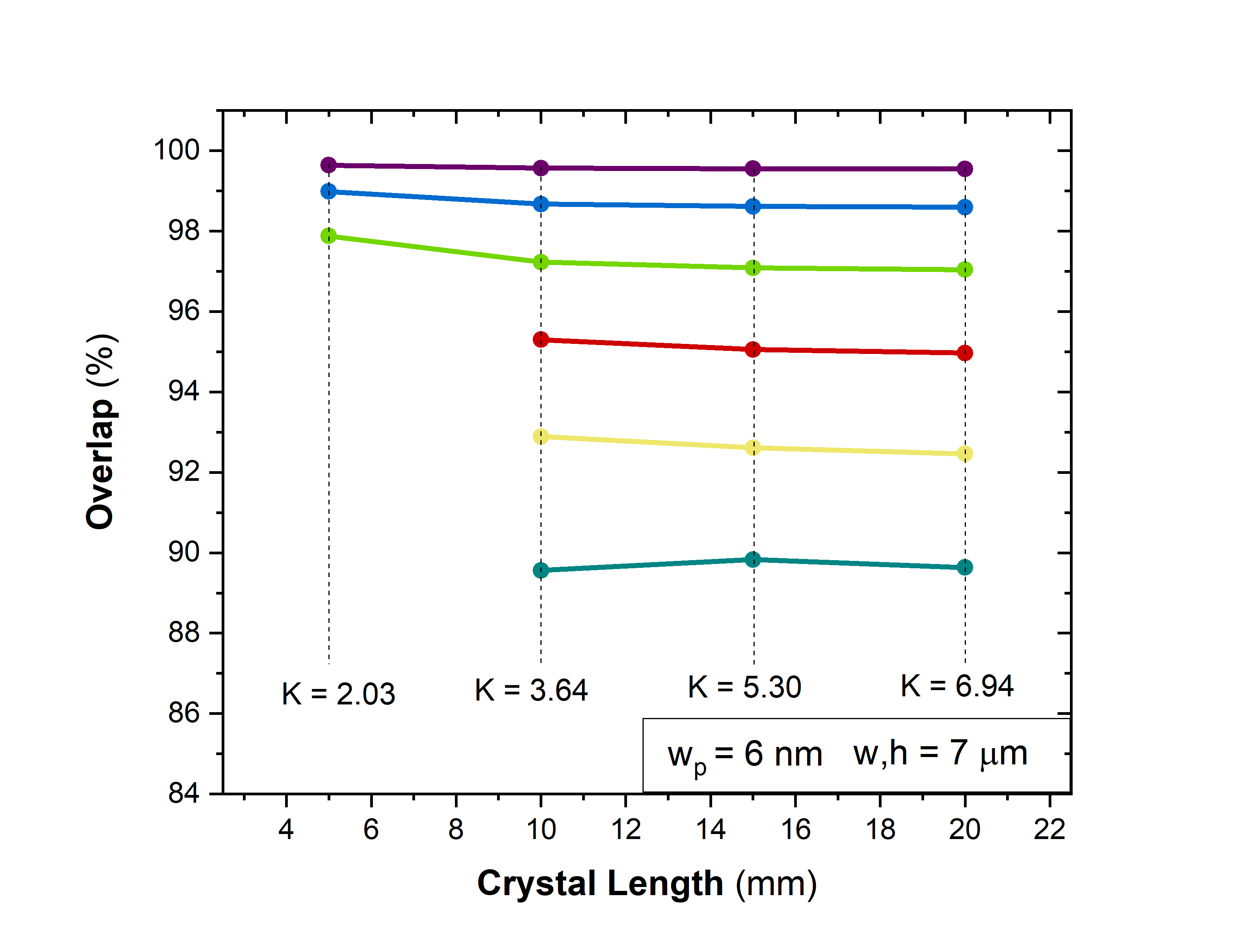}
\caption{Overlap of the first 6 Schmidt temporal modes as a function of crystal. Element 3.3 in Table \ref{Tabla}.}
\label{overlap_L}
\end{figure}

In a similar way, Fig.~\ref{overlap_L} shows that the waveguide length has no effect on the mode overlap. The Schmidt numbers, and hence effective mode numbers, for different lengths are marked in the plot. We remark that the real number of non zero modes (that we compute with the singular value decomposition of the JSA), is always higher than the Schmidt number $K$. For example, in Fig.~\ref{overlap_wp}, at a pump width of 4 nm, we have $K=2.30$ and 4 Schmidt modes, while when $K=3.54$, at a pump width of 6 nm, we already have 6 Schmidt modes. It is therefore worth noting that the Schmidt number $K$ is an indicative of the number of Schmidt modes present, but not their exact quantity \cite{Christ11}. In particular, if $m$ is equal to the real number of Schmidt modes, then $K\leq m$. These two figures correspond to the elements 3.1 and 3.3 in Table \ref{Tabla}.  

\begin{figure}
\centering
\includegraphics[width=0.5\textwidth]{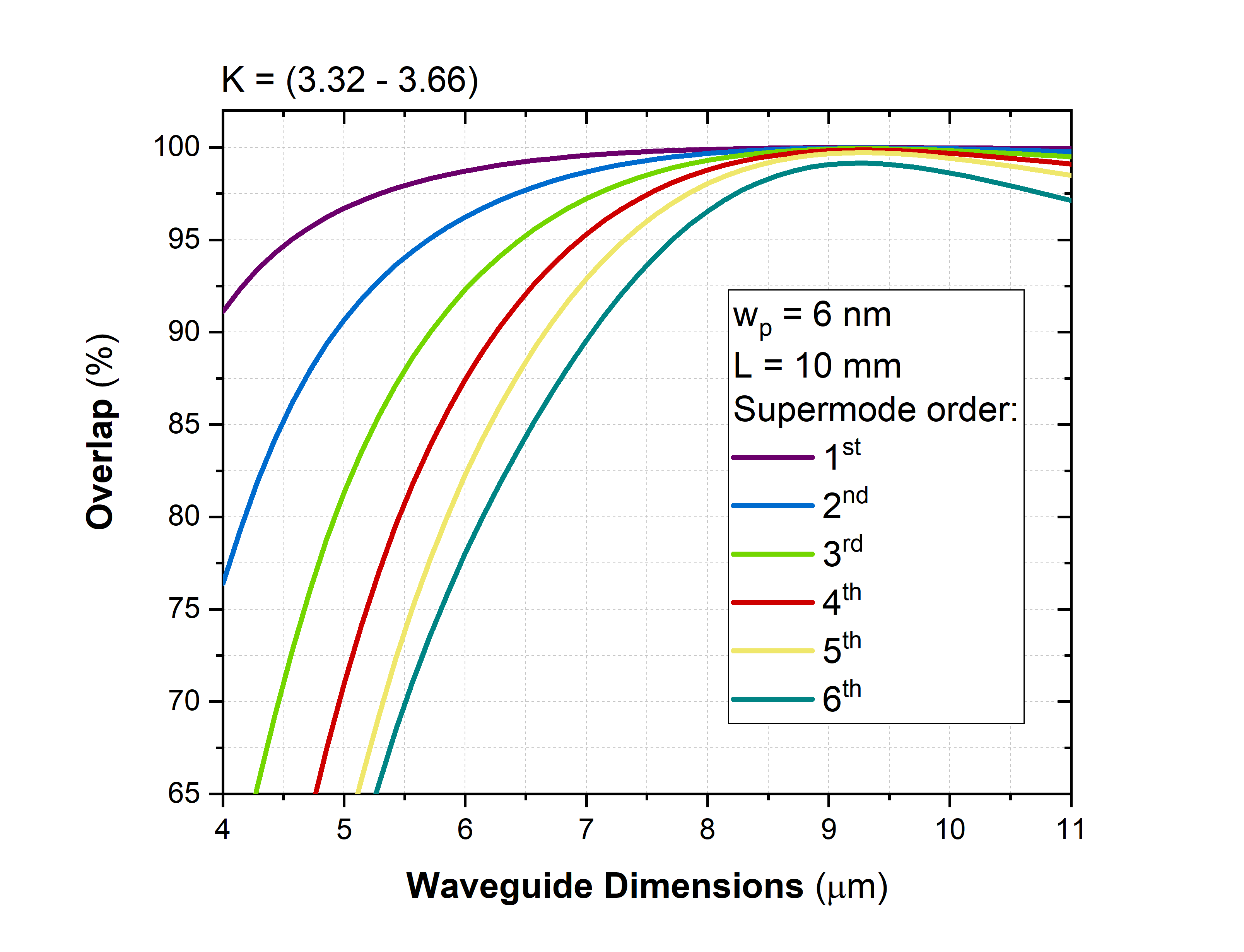}
\caption{Overlap of the first 6 Schmidt temporal modes as a function of waveguide dimensions. Element 3.2 in Table \ref{Tabla}.}
\label{overlap_wh}
\end{figure}
Completing Table \ref{Tabla}, Fig.~\ref{overlap_wh} shows the mode overlap as a function of the waveguide dimensions. We observe that the individual temporal mode overlaps have a common maximum at waveguide dimensions around 9 $\mu$m, as we computed in the SGVM study. Also, for every waveguide dimension, the similarity between modes decreases as the mode order increases (again the reason for this can be found in Appendix C). This dependence corresponds to element 3.2 in Table \ref{Tabla}.

\section{Conclusions}

In this work, we have studied the generation of multimode squeezed states in single-pass PDC in non-linear optical waveguides. Such states are appealing candidates for the realization of numerous quantum information applications, including measurement based quantum computation and multipartite quantum communication. For long distances and integrated application,  it is desirable to generate these states at telecom wavelengths. 

We have first set the theoretical grounds for the PDC process in a non-linear waveguide, leading to the generation of uncorrelated squeezed states, the ``supermodes'', and we identify them as a natural platform for continuous variable quantum information science. Then, we explored the different PDC configurations. 
 
In type \emph{0} and type \emph{I} PDC the generated signal and idler fields are indistinguishable. Nevertheless we have shown that these schemes are  not practical for CV protocols. Although highly multimode states are generated, the generated modes have indeed a too large spectrum to be accessible via homodyne detection.

Consequently, we targeted type \emph{II} PDC as the suitable scheme. In this case, the process generates  a number of two-mode squeezed states, that, upon interference on a beamsplitter, can be turned into the single-mode squeezed resource states. For this to work efficiently, signal and idler fields have to be spectrally indistinguishable. We showed that this requirement enforces the use of SGVM in PDC.

We then investigated the use of KTP, a nonlinear material which naturally provides SGVM for wavelengths around the telecommunications regime. We studied the number of generated squeezed states and their similarity as function of the PDC pump spectral width and waveguide length. In addition, we explored the influence of the waveguide dimensions on the generated PDC state. Waveguide dimensions are usually only set to ensure spatially single-mode operations. However, it turns out that they play a crucial role for achieving spectrally indistinguishable signal and idler fields at a desired wavelength. 

Our calculations reveal that a waveguide width and height of $9\,\mu$m facilitates SGVM at a signal and idler central wavelength of $1550\,$nm. In this configuration, both the waveguide length and pump spectral bandwidth can be used to tune the number of generated states, while having no detrimental impact on signal-idler similarity whatsoever. 

In conclusion we have explored symmetric group velocity  matching in non-linear waveguides for the tailored generation of scalable quantum resource for continuous variable based technologies.  Moreover we have identified the suited processes and materials along with realistic parameters for their generation at telecommunication wavelengths. 

\section*{Acknowledgments}
This  work  has  been  supported  by  the  European  Research Council under the Consolidator Grant COQCOoN (Grant No.  820079), by the QuantERA grant QuICHE, and by the German BMBF. VRR acknowledges support from the DGA.

\bibliography{biblioT}
\appendix{}
\section{Kleinman's Symmetry}\label{Klein}
When the nonlinear susceptibility tensor is frequency independent and the material is lossless, one can reduce the third rank tensor $\chi^{(2)}$ to a 3 by 6 matrix \cite{Boyd}:

\begin{equation}
    d_{il} = \frac{1}{2}\chi^{(2)}_{ijk}
\end{equation}

This matrix tells how the components of the second order non-linear polarization vector relates to the possible combinations of electric field components of the form $E_iE_j$. If an entry of this matrix is zero, it means that the non-linear process associated to the combination of those components does not contribute to the component of the polarization vector. If one works with linearly polarized fields where the polarization direction coincides with one of the defined material's optical axis, then the zero entries of $d_{il}$ give the corresponding not allowed processes.

For KTP, the $d$ matrix is given by \cite{indexrefraction}:
\begin{equation}
     d_{il} = \begin{pmatrix}
0 & 0 & 0 & 0 & -4.6 & 2.2 \\
-2.2 & 2.2 & 0 & -4.6 & 0 & 0 \\
-4.6 & -4.6 & 25 & 0 & 0 & 0 \\
\end{pmatrix}
\ \mathrm{pm/V}
\end{equation}
In our context, each of the non-zero entries can be seen as a type \emph{0}, type \emph{I} or type \emph{II} process, defined by the polarization (along $x$, $y$ or $z$) of the pump, signal and idler fields.

In this respect, along the computation in the paper we took the propagation direction (parallel to the waveguide) to be $x$ and the vertical polarization (parallel to the crystal cut) to be $z$. Hence, the vertical polarization is associated with the index of refraction $n_z$, and the horizontal polarization with the index $n_y$. The type \emph{II} process was described in terms of polarization as $(y\rightarrow z,y)$ (pump polarized in the $y$ direction, signal in the $z$ direction and idler also in the $y$ direction). The corresponding second order susceptibility tensor element is then $\chi^{(2)}_{232}$, which in the contracted notation of the matrix $d_{il}$ that corresponds to the element $d_{24}=-4.6$ pm/V. The fact that this element is non-zero tells us that the process is allowed for KTP.

We tried all possible combinations giving an allowed type \emph{II} process in KTP, finding that those combinations of polarizations were actually the only ones giving symmetric group velocity matching conditions at $1550$ nm, and therefore it is the one implicit in the manuscript. However, we remark that with other polarization combinations or other non-linear materials, different wavelengths fulfilling the symmetric group velocity matching condition could be found even for the interesting case of waveguide dimensions compatible with spatial single-mode propagation.

\section{Higher order contribution in the mismatch}\label{hig-ord}

We have seen that one of the conditions for being under SGVM is that the second and higher order terms of Eq.~(\ref{taylor_exp}) should be negligible compared to the first order contribution. As we are performing a two-variable Taylor expansion, the higher order contributions are expected to get bigger as we get further away from the central frequencies $\omega_s^0=\omega_i^0$.

Inserting Eq.~(\ref{index_Ref}) in Eq.~(\ref{wav_index}) we can numerically compute the wavevectors for pump, signal and idler, hence allowing the calculation (up to numerical error), of the wavector mismatch $\Delta k$,  by Eq.~(\ref{mismatch_qpm}), where the poling period $\Lambda$ would be computed beforehand for phasematching the central frequencies.

The first order term, that we will denote here as $F(\omega_s,\omega_i)$, can be calculated as:
\begin{equation}
    F(\omega_s,\omega_i) = \gamma_s(\omega_s-\omega_p^0/2) + \gamma_i(\omega_i-\omega_p^0/2);
\end{equation}
where the coefficients have been defined in Eq.~(\ref{taylor_coeff}).  

Therefore, the higher order terms, denoted as $O(\omega_s,\omega_i)$ can be computed by:
\begin{equation}
    O(\omega_s,\omega_i) = \Delta k(\omega_s,\omega_i) -  F(\omega_s,\omega_i)
\end{equation}
In general, the higher order contributions depend on waveguide dimensions and temperature. In our case, we are interested in seeing the contribution of these terms under the phasematching conditions, \emph{i.e.}, when $\gamma_s=-\gamma_i$, so that our waveguide dimensions are fixed in order to satisfy the condition.  

As we have seen in the paper, temperature does not play an important role in the wavelength where the first order term provides the phasematching condition. We have checked that temperature does not change appreciably the relative values of the higher order terms with respect to the first order term either.

Also, we will show here only the case of KTP, although the same procedure could be performed for any non-linear material.
\begin{figure}[H]
\centering
\includegraphics[width=0.5\textwidth]{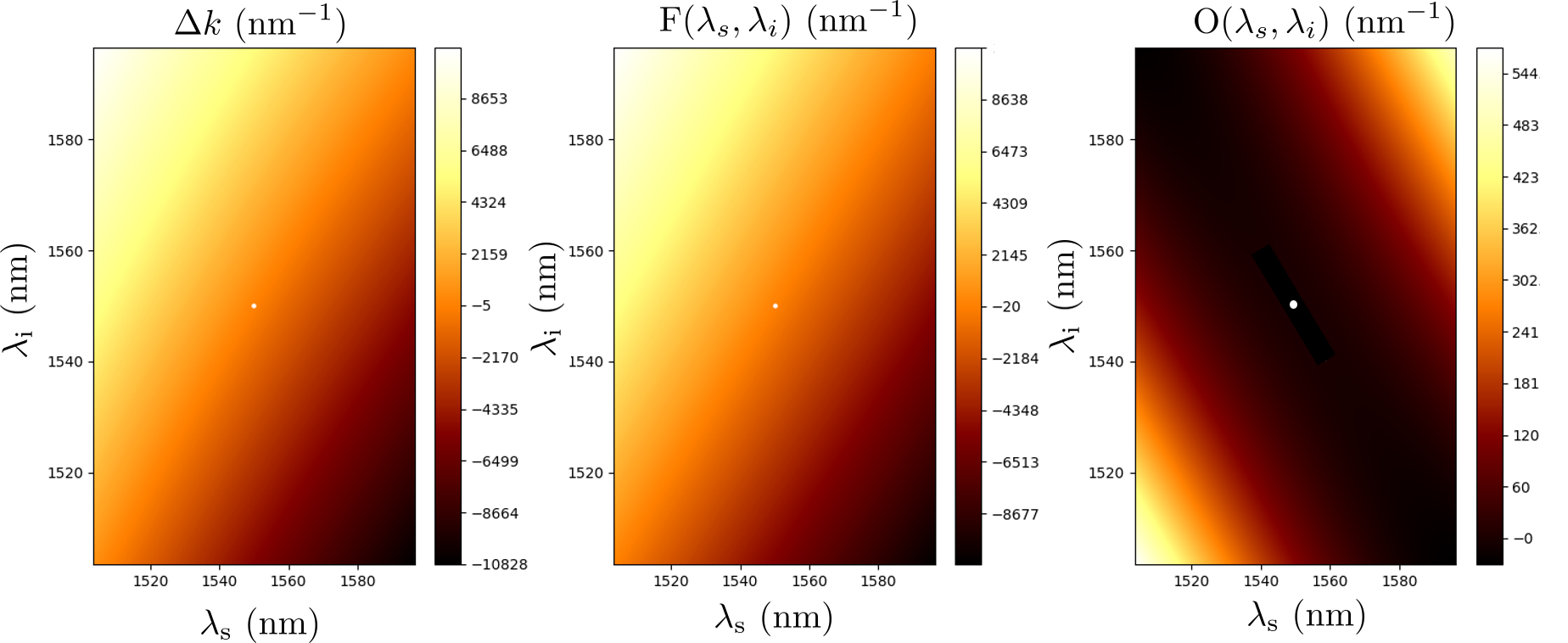}
\caption{Left to right: numerical mismatch, $\Delta k(\lambda_s,\lambda_i)$, first order term, $F(\lambda_s,\lambda_i)$ and higher order terms, $O(\lambda_s,\lambda_i)$ as a function of signal and idler wavelengths, $\lambda_s,\lambda_i$.}
\label{higher_order_mismatch}
\end{figure}

Fig.~\ref{higher_order_mismatch} shows the three functions $\Delta k$, $F$ and $O = \Delta k - F$ under phasematching conditions for KTP as a function of the signal and idler wavelengths (recall that $\lambda = 2\pi/\omega$). We can see that the mismatch function is effectively well described by the first order term in the wavelength region considered (that is bigger than the Schmidt modes bandwidth calculated in the paper). 

We therefore conclude that, for KTP, and under our configuration, the higher order terms in the mismatch can be discarded in practice and the SGVM condition holds.

\subsection{Analitical mode overlap calculation}

In Fig.\ref{overlap_wp}, Fig.\ref{overlap_L} and Fig.\ref{overlap_wh}, we can see that the overlap between the modes decreases as the mode order increases.

The reason for this is two-folded: first, as the mode order increases, their spectral width does too, reaching frequencies further away from the central frequency, where the Taylor expansion of Eq.(\ref{taylor_exp}) is less precise and hence the modes are expected to present more differences between them. Their overlap is expected to be smaller, as we confirm.

The second reason resides in the mathematical structure of the modes itself. If we approximate the phasematching function (sinc function), by a gaussian function, the Schmidt modes can be calculated analytically, giving two sets of Hermite-Gauss modes of the form:
\begin{equation}
    \mathrm{HG}_n(x) = \frac{1}{\sqrt{n!\sqrt{\pi}2^nw}}H_n(x)\exp(-(x-x_0)^2/2w^2)
    \label{hermite-gauss}
\end{equation}
where $H_n(x)$ is the n-th order Hermite polynomial and $w$ is the associated width of the zeroth-order function, which is a gaussian. Our Schmidt modes, depicted in Fig.\ref{waveguide_scheme}, are not Hermite-Gauss, but they approximate them fairly well.

If we try to compute the n-th order overlap defined in Eq.(\ref{overlap}) from two sets of Hermite-Gauss modes with different widths, $w$ and $w'$, we find, using Eq.(\ref{hermite-gauss}):
\begin{equation}
\begin{aligned}
    o_n &= \frac{\int_{-\infty}^{\infty}H^2_n(x)e^{-a_1x^2+b_1x}\mathrm{d}x}{\sqrt{\int_{-\infty}^{\infty}H^2_n(x)e^{-a_2x^2+b_2x}\mathrm{d}x\int_{-\infty}^{\infty}H^2_n(x)e^{-a_3x^2+b_3x}\mathrm{d}x}}\\
    a_1 &= \frac{1}{2}\left(\frac{1}{w^2}+\frac{1}{w'^2}\right)\qquad
    b_1 = x_0\left(\frac{1}{w^2}+\frac{1}{w'^2}\right)\\
    a_2 &= \frac{1}{w^2}\qquad\qquad\qquad\quad
    b_2 = \frac{2x_0}{w^2}\\
    a_3 &= \frac{1}{w'^2}\qquad\qquad\qquad\quad
    b_3 = \frac{2x_0}{w'^2}\\
    \label{overlap_analitico}
    \end{aligned}
\end{equation}
and therefore the problem is reduced to finding the integral:
\begin{equation}
    L \equiv \int_{-\infty}^{\infty}H^2_n(x)e^{-ax^2+bx}\mathrm{d}x
    \label{integral}
\end{equation}
for the solution of this integral one can define both the whole family of two variable Hermite polynomials $H_n(x,y)$ and the family of two-indices Hermite polynomials $H_{n,m}(x,y,w,z|\tau)$ \cite{Hermiteintegrals}. By the method of the generating function, one finds that the solution to the integral $L$ in Eq.(\ref{integral}) is an evaluation of the two-index (with the same index $n$ and $n$) Hermite polynomial in specific coordinates depending only on $a$ and $b$.

\begin{figure}[H]
\centering
\includegraphics[width=0.5\textwidth]{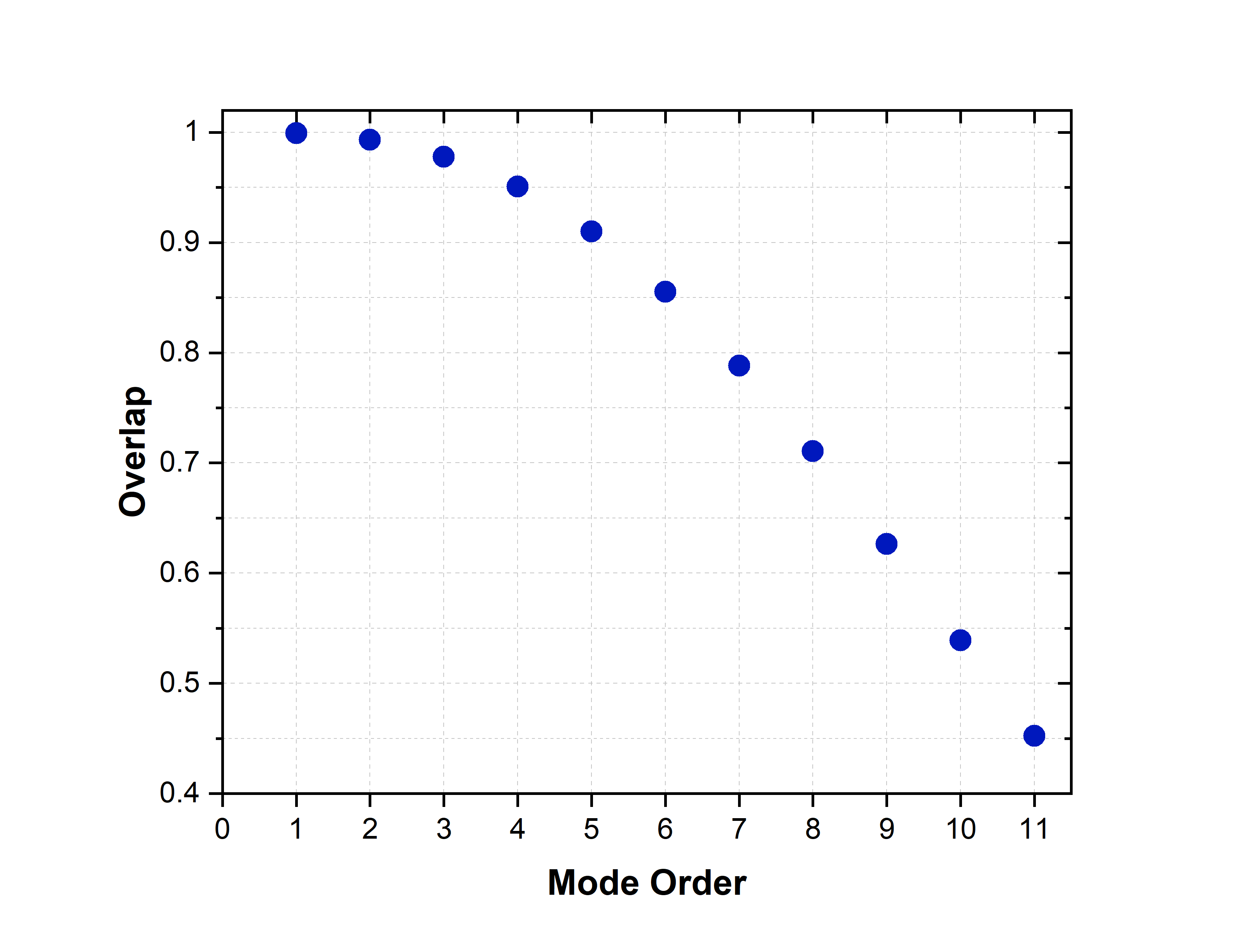}
\caption{Schmidt modes overlap as a function of the mode order for two sets of Hermite-Gauss modes with relative widths of $w/w'=0.95$.}
\label{analitical_o}
\end{figure}

For typical values of relative widths for signal and idler Schmidt modes computed in our simulations, the overlap between modes decreases slightly and linearly. If we also model the fact that our modes are expected to be less similar as the mode order increases, due to their larger spectral width, by making their relative widths to increase with the mode order, then we obtain  Fig.{\ref{analitical_o}}. In this Figure, the relative widths for the firsts modes (gaussians) is set to $w/w' = 0.95$. From this value, we decrease their relative widths by 2\% for every higher mode order. We observe a decreasing overlap function with the mode order, in very good accordance with the results from the manuscript. We conclude that both the dispersion and the mathematical nature of the modes are responsible for the observed behaviour of the overlap as the mode order increases.

\end{document}